\documentclass[a4paper,sort&compress,oneside,preprint,12pt,number]{elsarticle}
\usepackage[utf8]{inputenc}
\usepackage[T1]{fontenc}
\usepackage{color}
\usepackage{amssymb,amsmath,amsthm,mathtools}
\usepackage{subfigure,overpic}
\usepackage{hyperref}
\usepackage{overpic}

\usepackage[left=2cm,right=2cm,top=2cm,bottom=2cm]{geometry}

\newcommand{\ii}{\mathrm{i}}
\DeclareMathOperator{\arccosh}{arccosh}
\DeclareMathOperator{\arcsinh}{arcsinh}

\usepackage{thmtools}
\declaretheoremstyle[headfont=\kpfonts]{normalhead}

\newtheorem{theorem}{Theorem}[section]

\theoremstyle{definition}

\journal{?}

\begin{document}

\begin{frontmatter}

\title{On the proximity of Ablowitz-Ladik and discrete Nonlinear
Schr\"odinger models: A theoretical and numerical study of Kuznetsov-Ma
solutions}

\affiliation[label1]{organization={Mathematics Department, %
             California Polytechnic State University},
             city={San Luis Obispo},
             postcode={93407-0403},
             state={CA},
             country={USA}}

\affiliation[label12]{organization={Department of Mathematics and Statistics, %
             and Computational Science Research Center,
             San Diego State University},
             city={San Diego},
             postcode={92182-7720},
             state={CA},
             country={USA}}

\affiliation[label2]{organization={Department of Mathematics, %
             University of Kansas},
             city={Lawrence},
             postcode={66045},
             state={KS},
             country={USA}}

\affiliation[label3]{organization={Grupo de F\'{i}sica No Lineal, Departamento de F\'{i}sica Aplicada I,
Universidad de Sevilla. Escuela Polit\'{e}cnica Superior, C/ Virgen de \'{A}frica, 7, 41011-Sevilla, Spain\\
Instituto de Matem\'{a}ticas de la Universidad de Sevilla (IMUS). Edificio Celestino Mutis. Avda. Reina
Mercedes s/n, 41012-Sevilla, Spain}}

\affiliation[label4]{organization={Department of Mathematics and Statistics, %
             University of Massachusetts Amherst},
             city={Amherst},
             postcode={01003-4515},
             state={MA},
             country={USA}}

\affiliation[label5]{organization={Department of Mathematics, %
             University of Thessaly},
             city={Lamia},
             postcode={35100},
             country={Greece}}

\author[label1]{Madison L. Lytle}
\ead{mllytle@calpoly.edu}
\author[label1,label12]{Efstathios G. Charalampidis}
\ead{echaralampidis@sdsu.edu}
\author[label2]{Dionyssios Mantzavinos}
\ead{mantzavinos@ku.edu}
\author[label3]{Jesus Cuevas-Maraver}
\ead{jcuevas@us.es}
\author[label4]{Panayotis G. Kevrekidis}
\ead{kevrekid@umass.edu}
\author[label5]{Nikos I. Karachalios}
\ead{karan@uth.gr}

\begin{abstract}
In this work, we investigate the formation of time-periodic solutions with a non-zero background
that emulate rogue waves, known as Kuzentsov-Ma (KM) breathers, in physically relevant
lattice nonlinear dynamical systems.~Starting from the completely integrable Ablowitz-Ladik
(AL) model, we demonstrate that the evolution of KM initial data is proximal to that of
the non-integrable discrete Nonlinear Schr\"odinger (DNLS) equation for certain parameter
values of the background amplitude and breather frequency.~This finding prompts
us to investigate the distance (in certain norms) between the evolved solutions of both
models, for which we rigorously derive and numerically confirm an upper bound.~Finally, 
our studies are complemented by a two-parameter (background amplitude and frequency) bifurcation analysis of numerically exact,
KM-type breather solutions to the DNLS equation.~Alongside
the stability analysis of these waveforms reported herein, this work additionally showcases
potential parameter regimes where such waveforms with a flat background may emerge in the DNLS setting.
\end{abstract}


% \begin{highlights}
% \end{highlights}


\end{frontmatter}

\section{Introduction and Motivation}

The study of dispersive nonlinear lattice dynamical
models has been a topic of considerable interest over the
past decades~\cite{Aubry2006,Flach2008}. Among the 
physical areas motivating the relevant developments,
one can single out the particular contributions from the study of optical waveguides~\cite{LEDERER20081}
(but also continuum photorefractive media with periodic
potentials) and the exploration of mean-field
atomic Bose-Einstein condensates (BECs) in the presence
of periodic external (so-called optical lattice) 
potentials~\cite{RevModPhys.78.179}. However, 
relevant models, computations, and experiments
are by no means limited to these subfields;
rather, they broadly extend to other contexts, including, among
others, nonlinear variants of electrical circuits~\cite{remoissenet},
elastically interacting beads within granular crystal
metamaterials~\cite{Nester2001,yuli_book,Chong2018},
superconducting Josephson junction arrays~\cite{alex,alex2},
micromechanical arrays of cantilevers~\cite{cantilevers},
and DNA denaturation models~\cite{peybi}.

There has been a considerable wealth of models relevant
to these spatially discrete applications, including
Klein-Gordon and Fermi-Pasta-Ulam-Tsingou~\cite{imamat,FPUreview}.%
~However, the most universal dispersive lattice nonlinear model
is, arguably, the discrete nonlinear 
Schr{\"o}dinger (DNLS) equation~\cite{kev09,chriseil}.
It serves as the prototypical vehicle for the study
of solitary waves, instabilities, and dynamics in discrete
nonlinear optics~\cite{cole} while also holding considerable relevance in atomic BECs~\cite{RevModPhys.78.179}.
To this day, it continues to play a substantial role in 
modern developments concerning, e.g., topological
lattices~\cite{cole}, flat bands~\cite{leykam2018artificial,DanieliAndreanovLeykamFlach+2024+3925+3944},
and many others. 
% Additionally, it has the particularly
% interesting feature from a mathematical physics perspective
% of the existence of an integrable analogue thereof~\cite{AblowitzPrinariTrubatch}.%
From a mathematical physics perspective, it has an additional, particularly
interesting feature: the existence of an integrable analog ~\cite{AblowitzPrinariTrubatch}.%
% ~This, in turn, creates the potential for numerous studies of breaking of
% integrability, and consideration of effects of perturbations on conservation
% laws, solitary features etc. in suitable interpolations between the integrable
% and non-integrable variants of the model~\cite{salerno1992quantum}.
~This, in turn, creates the potential for numerous studies concerning the breaking of
integrability, perturbation effects on conservation
laws, solitary features, etc. in suitable interpolations between the integrable
and non-integrable variants of the model~\cite{salerno1992quantum}.

An aspect of DNLS solitary waves that has been
of interest in recent years is its potential
to feature large-amplitude rogue or freak waves. 
The study of such waves~\cite{Kharif2009} has recently attracted
considerable attention due to the emergence
of experimental capabilities that enable the detection
and visualization of these patterns at the level of
nonlinear waves in optical systems~\cite{Kibler2010,Kibler2012,DeVore2013,Frisquet2016},
fluid settings in water tanks~\cite{Chabchoub2011,Chabchoub2012,Chabchoub2014},  
% and other areas such as 
plasmas~\cite{Bailung2011}, 
and even BECs~\cite{romeroros}. These developments
have been summarized in a wide range of
reviews, including~\cite{Yan2012a,Onorato2013,Dudley2014,Mihalache2017,natrevphys}. In the discrete realm, the work of~\cite{sotoc} 
% planted the seed by showcasing 
showcased
the existence of all the central
nonlinear waveforms in the integrable Ablowitz-Ladik (AL)
model: i.e., the Peregrine soliton (P)~\cite{Peregrine1983},
the
Akhmediev breather (AB)~\cite{Akhmediev1986}, as well
as the Kuznetsov-Ma (KM) soliton~\cite{Kuznetsov1977,Ma1979}.

% Ever since the emergence of such patterns in discrete integrable
% media, the question has been raised of their potential
% observability~\cite{maluck,HOFFMANN20183064}.
The emergence of such patterns in discrete integrable
media raised the question of their potential
observability~\cite{maluck,HOFFMANN20183064}.
% The expectation
% from these works has been that considering models interpolating
% between the integrable (AL) and non-integrable (DNLS) limit,
% extreme events are more likely to exist in the former, rather
% than in the latter. 
When considering models interpolating
between the integrable (AL) and non-integrable (DNLS) limit,
these works set the expectation that
extreme events are more likely to exist in the former, rather
than in the latter. 
% On the other hand, in recent years,
% some of the present authors have been exploring the
% proximity between the DNLS and the AL model at the level
% of analytical estimates~\cite{JDE0,JDE2} and the
% continuation between the latter and the former
% at the level of numerical computations~\cite{sullivan}.
On the other hand, in recent years,
some of the present authors have explored the
proximity between the DNLS and the AL model through analytical estimates~\cite{JDE0,JDE2} as well as
continuations between the two
via numerical computations~\cite{sullivan}.
% ~It is the scope
% of the present work to bring these different elements
% of the literature together, operating at the nexus of
% extreme solutions of lattice nonlinear dynamical systems 
% of both the integrable (AL) and the non-integrable (DNLS) kind,
% and exploring the potential continuation of KM solutions from 
% the former towards the latter.
~The present work aims to bring these different elements
of the literature together, operating at the nexus of
extreme solutions of lattice nonlinear dynamical systems 
of both the integrable (AL) and the non-integrable (DNLS) kind,
and exploring the potential continuation of KM solutions from 
the former towards the latter.
~Moreover, the present work %and considering 
considers such questions
from the complementary perspectives of rigorous estimates,
as well as of numerical computations involving the existence,
spectral stability, and nonlinear dynamics of such states. 
We find that the relevant states can often
be continued and we can identify corresponding waveforms
in the DNLS limit. {This is corroborated by the 
presence of the analytical estimates.}

Our presentation is structured
as follows.~In Section 2, we present the model
setup and relevant solutions of interest.
In Section 3, we analyze some preliminary
numerical computations showcasing the proximity
over long times (for suitable solution parameters)
of the KM waveforms for the AL and the DNLS models.
This finding then motivates Section 4, % us in Section 4 to provide
where we provide 
a rigorous analysis of the growth over time
of the distance between the solutions of
the DNLS and the AL systems in the case of non-zero
boundary conditions. We establish that this growth
is linear in time with a prefactor 
whose dependence on the size of initial data is
analyzed.
%that is dependent
%on the distance between the two models' initial %conditions.
In Section 5 we present a continuation of the KM branch of
solutions of the DNLS model, along with a discussion of the spectral stability of the pertinent solutions.~The work is summarized and our
conclusions are presented in Section 6.

\section{The model setup}\label{sec_1}
% ======================================================================
% In our theoretical and computational analysis that follow next,
In our subsequent theoretical and computational analysis,
we consider
the Salerno model~\cite{salerno1992quantum}, explicitly given by
\begin{equation}
\ii\dot{\Psi}_{n}+C(\Delta_d\Psi)_n%
+g\left(\Psi_{n+1}+\Psi_{n-1}\right)|\Psi_{n}|^{2}+%
2\left(1-g\right)|\Psi_{n}|^{2}\Psi_{n}=0,
\label{salerno0}
\end{equation}
that interpolates between the completely integrable Ablowitz-Ladik (AL)
model~\cite{ablowitz_ladik_1975,ablowitz_ladik_1976} (at $g=1$) and the
discrete Nonlinear Schr\"odinger (DNLS) equation~\cite{kev09} (at $g=0$).%
~In Eq.~\eqref{salerno0}, $(\Delta_d\Psi)_n=\Psi_{n+1}-2\Psi_{n}+\Psi_{n-1}$,
and the coupling constant $C$ between nearest neighboring sites is given by
$C=1/h^{2}$ where $h$ is the lattice spacing between adjacent sites.

% Our main interest is the study of breather solutions on a non-zero background
% of Eq.~\eqref{salerno0}.
Our main interest is the study of breather solutions of Eq.~\eqref{salerno0} on a non-zero background.
~To that end, we consider the separation of variables ansatz:
\begin{equation}
\Psi_{n}=\psi_{n}\,e^{2iq^{2}t}, \quad q>0,
\label{eq:sep_vars}
\end{equation}
where $q$ represents the background amplitude of the solution of interest,
and thus $2q^{2}$ is the oscillation frequency of the background.~Upon inserting
Eq.~\eqref{eq:sep_vars} to Eq.~\eqref{salerno0}, we obtain:
\begin{equation}
\ii\dot{\psi}_{n}+C(\Delta_d\psi)_n%
+g\left(\psi_{n+1}+\psi_{n-1}\right)|\psi_{n}|^{2}+%
2\left[\left(1-g\right)|\psi_{n}|^{2}-q^{2}\right]\psi_{n}=0,
\label{salerno}
\end{equation}
which is the principal model equation of interest for studying time-periodic
solutions $\psi_{n}(t)=\psi_{n}(t+T_{b})$ of period $T_{b}=2\pi/\omega_{b}$ and
frequency $\omega_{b}$.~For $C=1$, the AL model ($g=1$) admits explicit rogue
wave solutions~\cite{sotoc} including the (discrete) Kuznetsov-Ma (KM), Peregrine (P),
and (spatially-periodic) Akhmediev breathers (AB).~In particular, the KM breather to
the AL model is explicitly given by:
\begin{equation}
\psi_{n}(t)=q\frac{\cos{\left(\omega_{b} t+\ii\theta\right)+G\cosh{\left(r n\right)}}}%
{\cos{\left(\omega_{b} t\right)+G\cosh{\left(r n\right)}}},
\label{km_exact}
\end{equation}
with parameters $\theta=-\arcsinh{\left(\omega_{b}/(2q^2)\right)}$, $m=(1+q^2)/q^2$,
$r=\arccosh{\left[\left(\cosh{\theta}+m-1\right)/m\right]}$, and
$G=-\omega_{b}/\left(2q^2\sqrt{m}\sinh{r}\right)$.~Note that the Peregrine
breather~\cite{sotoc,Peregrine1983} is the limiting case $T_{b}\mapsto\infty$
(or $\omega_{b}\mapsto 0$) of the KM solution of Eq.~\eqref{km_exact}.~In this work, we solely
focus on studies revolving around KM-type solutions to both the AL ($g=1$) and DNLS
($g=0$) models.~Hereafter, we fix $C=1$ corresponding to a lattice with
unit spacing, i.e., $h=1$.

\section{Preliminary Numerical Studies}\label{sec_2}
% ======================================================================
Recently, in~\cite{sullivan}, the homotopic continuation of KM breathers (and
their stability) in the Salerno model was considered, and numerical results
showcased the emergence of KM-type breathers. Some
of these KM-type solutions could be continued all the way to the DNLS limit (i.e., $g=0$
in Eq.~\eqref{salerno0}), but featured an oscillatory background.~Attempts for identifying
such solutions on a flat background by starting from the anti-continuum limit
(i.e., $C=0$) are reported in~\cite{antibreathers}, although the relevant waveforms
obtained do not enjoy rogue-like behavior.~That is to say, the waveforms therein do
not seem to ``appear out of nowhere and disappear without a trace''~\cite{AKHMEDIEV2009675}.

Herein, we take a different path and examine the possibility of identifying
KM solutions to the DNLS by considering the recent advances on the closeness
of localized structures between the AL and DNLS models~\cite{JDE2,HENNIG2022346,JDE1}.%
~{In~\cite{HENNIG2022346}}, %that work, 
the proximity of Peregrine waveforms in the AL and DNLS models
was examined, showing their persistence in DNLS models when $q$, i.e.,
the value of the background, is small.~Motivated by this finding, we now explore
numerically whether KM breathers with small background amplitudes $q\ll1$ and frequencies
$\omega_{b}\ll1$, i.e., close to the Peregrine limit, may persist in the DNLS
model.~In our computations that follow in this section, we utilize a lattice
$[-N/2,N/2]$ consisting of $N=600$ nodes (unless stated otherwise), where periodic
boundary conditions are supplemented in Eq.~\eqref{salerno}. %***

Then, the initial-value problem (IVP) consisting of Eqs.~\eqref{salerno}-\eqref{km_exact} is
solved numerically by using MATLAB's built-in Adams-Bashforth-Moulton \textsc{ode113}
solver (with relative and absolute tolerances $10^{-13}$).~We checked our numerical
results by additionally using the Bulirsch-Stoer method~\cite{hairer_1993} (with
same relative and absolute tolerances), and the results matched exactly.~The fidelity
of our computations utilizing both numerical methods was further checked by monitoring
the conserved quantities for the AL and DNLS models, respectively given by
\begin{align}
P_{\mathrm{AL}}(t)=\sum_{n}\log{\left(1+|\psi_{n}|^{2}\right)},
\label{eq:consv_AL}
\end{align}
and
\begin{align}
P_{\mathrm{DNLS}}(t)=\sum_{n} |\psi_{n}|^{2}.
\label{eq:consv_DNLS}
\end{align}
We found that the maximum relative errors, i.e.,
$\mathrm{max}\left(|P_{\mathrm{AL}}(t)-P_{\mathrm{AL}}(0)|/P_{\mathrm{AL}}(0)\right)$
(and similarly for the DNLS) were $\sim 10^{-14}$.

\begin{figure}[pt]
{\hskip -1.7cm
\begin{overpic}[height=.18\textheight, angle =0]{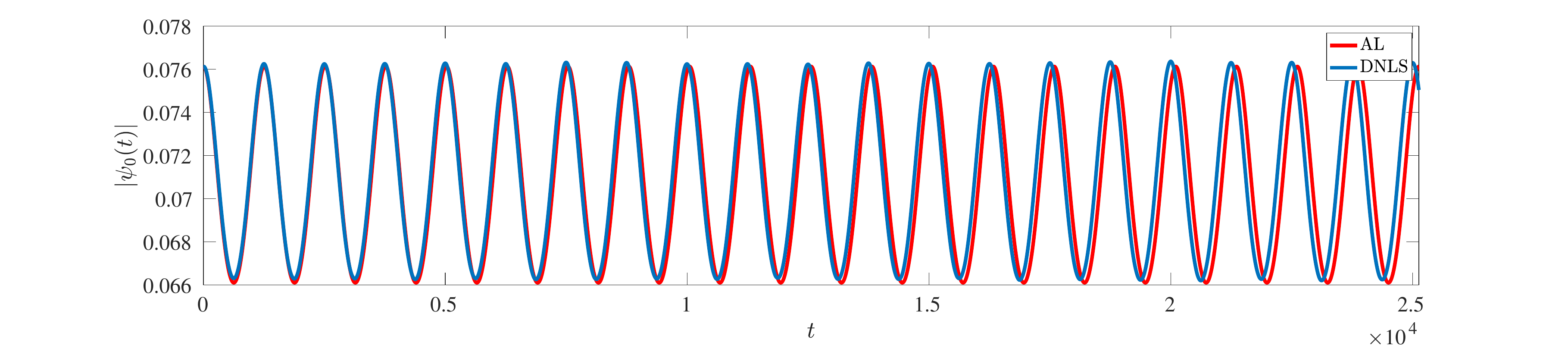}
\put(13.5,18.5){{\small (a)}}
\end{overpic}}\\
\hspace*{2.2cm}
\begin{overpic}[height=.18\textheight, angle =0]{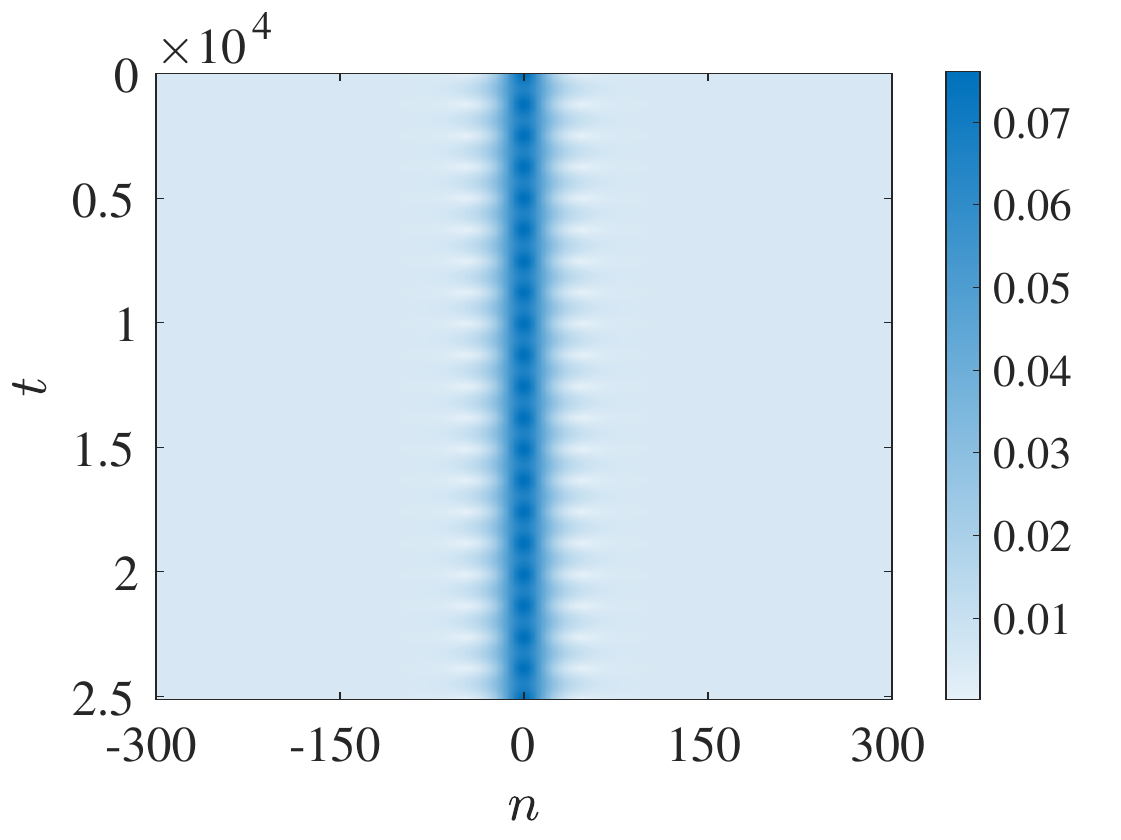}
\put(17,61){{\small (b)}}
\end{overpic}
\begin{overpic}[height=.18\textheight, angle =0]{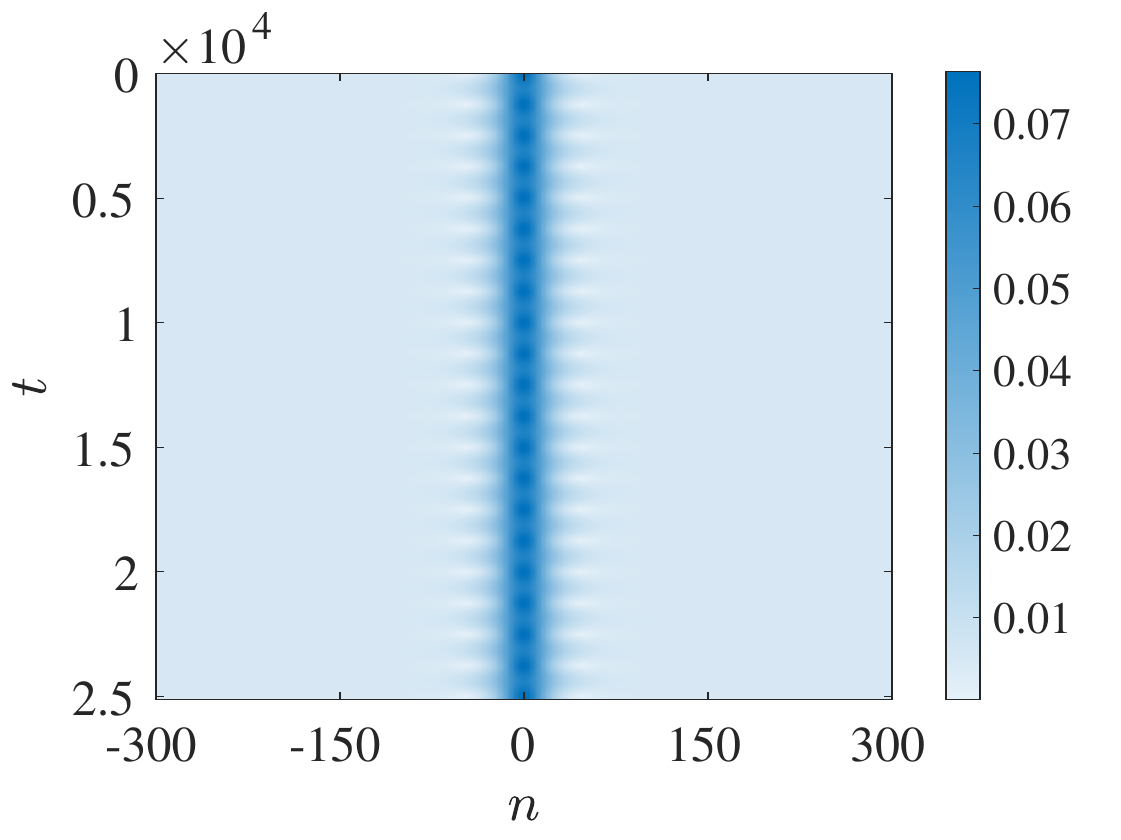}
\put(17,61){{\small (c)}}
\end{overpic}
\caption{(Color online) The spatio-temporal evolution of KM breathers in the AL
and DNLS models [cf.~Eq.~\eqref{salerno} for $g=1$ and $g=0$, respectively] with
$q=\omega_{b}=0.005$ in Eq.~\eqref{km_exact}.~Panel (a) compares the evolution of
the amplitude at the $n=0$ site, i.e., $|\psi_{0}(t)|$ with solid red (AL) and blue
(DNLS) lines (see, the legend therein).~Panels (b) and (c) showcase the spatio-temporal
evolution of the amplitude $|\psi_{n}(t)|$ of the KM solution in the AL and DNLS models,
respectively.~A lattice of $N=600$ sites was used with periodic boundary conditions.
\label{prelim_numer}}
\end{figure}

Let us now turn our focus to the spatio-temporal evolution of KM breathers
[cf.~Eq.~\eqref{km_exact}] in Eq.~\eqref{salerno} (again with $C=1$) for $g=1$
(AL) and $g=0$ (DNLS).~We use the analytical KM solution of Eq.~\eqref{km_exact} with
$\omega_{b}=0.005$ and $q=0.005$ as an initial condition for both models at $t=0$, and advance
Eq.~\eqref{salerno} forward in time for up to 20 periods, i.e.,
$t=20\times T_{b}\approx 2.5133\times10^{4}$.~Our results for these cases are summarized
in Fig.~\ref{prelim_numer}.~In particular, the panel (a) of the figure depicts the
temporal evolution of the amplitude at the $n=0$ site, i.e., $|\psi_{0}(t)|$
for the AL and DNLS models (i.e., $g=1$ and $g=0$ in Eq.~\eqref{salerno}).~It
can be discerned from this panel that the evolutions of the amplitudes are
close 
%enough
for a few periods of the time integration although a phase
lag in their evolution gradually gets manifested, and progressively becomes more and more
pronounced for longer times.~We complement this first round of numerical simulations
in panels (b) and (c) of Fig.~\ref{prelim_numer} where we summarize the spatio-temporal
evolution of the amplitude $|\psi_{n}(t)|$ of the solutions for the AL and DNLS models,
respectively.~Despite this phase lag reported in panel (a), panels (b) and (c) exhibit
similar qualitative (and quantitarive) features in the evolution of the waveforms.~

These findings now beg the question whether such a KM solution can be identified
numerically as a fixed point, i.e., as a numerically exact periodic orbit for the DNLS
case.~To explore this possibility, we followed the numerical approach of~\cite{sullivan,antibreathers}
%where 
{in which} a $T_{b}=2\pi/\omega_{b}$ time-periodic solution is sought by considering
the Fourier decomposition
\begin{align}
\psi_{n}=\sum_{m=-\infty}^{\infty}{\cal Z}_{n,m}e^{\ii m\omega_{b}t},
\label{eq:fourier_decomp}
\end{align}
where ${\cal Z}_{n,m}$ are the Fourier coefficients and $m$ are the Fourier
modes in time.~Then, upon plugging Eq.~\eqref{eq:fourier_decomp} into Eq.~\eqref{salerno}
we obtain a root-finding problem for ${\cal Z}_{n,m}$ (see, \cite{sullivan} for the
explicit form of the relevant root-finding problem), that is solved with high accuracy
by means of Newton's method.~We employ stopping criteria in the nonlinear residual and
successive iterates of $10^{-14}$.~It should be noted that the infinite sum of Eq.~\eqref{eq:fourier_decomp}
is truncated according to $|m|\leq k$ where $k=41$, and thus $2k+1=83$ Fourier modes
were employed.

\begin{figure}[pt]
\hskip -0.8cm
\begin{overpic}[height=.177\textheight, angle =0]{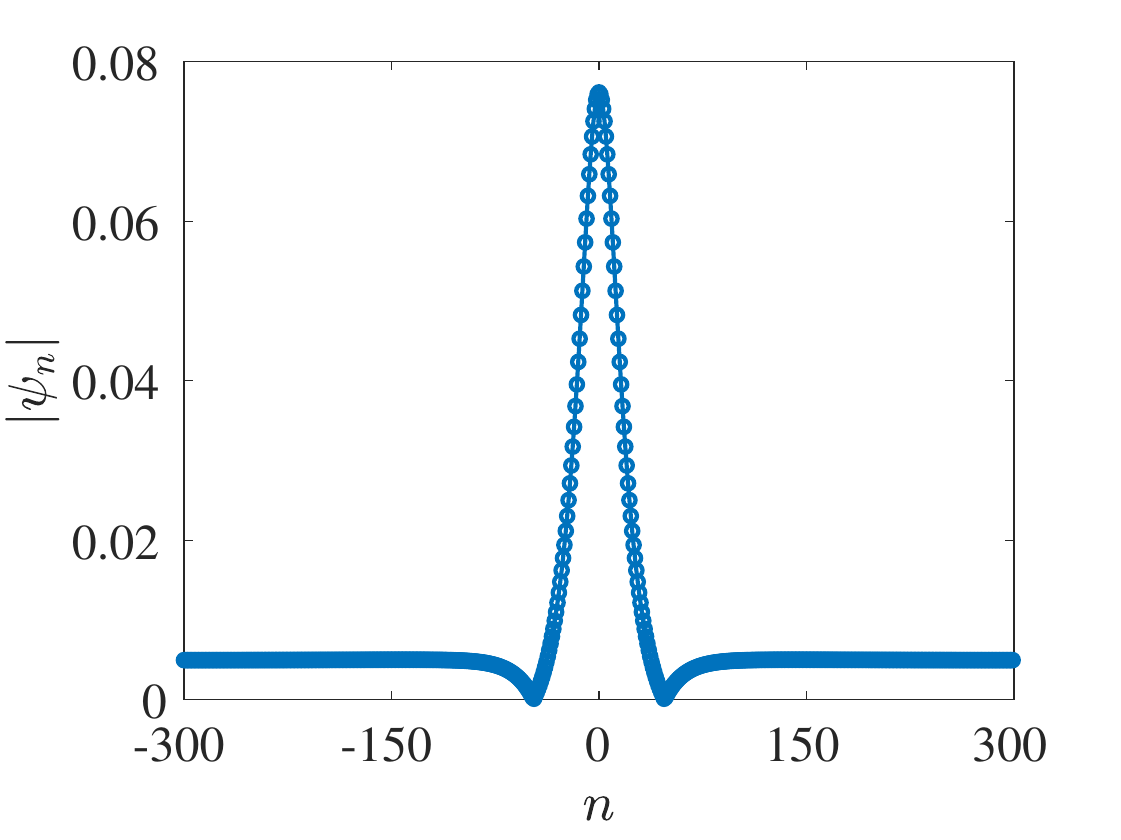}
\put(19,61){{\small (a)}}
\end{overpic}
\hskip -0.2cm
\begin{overpic}[height=.177\textheight, angle =0]{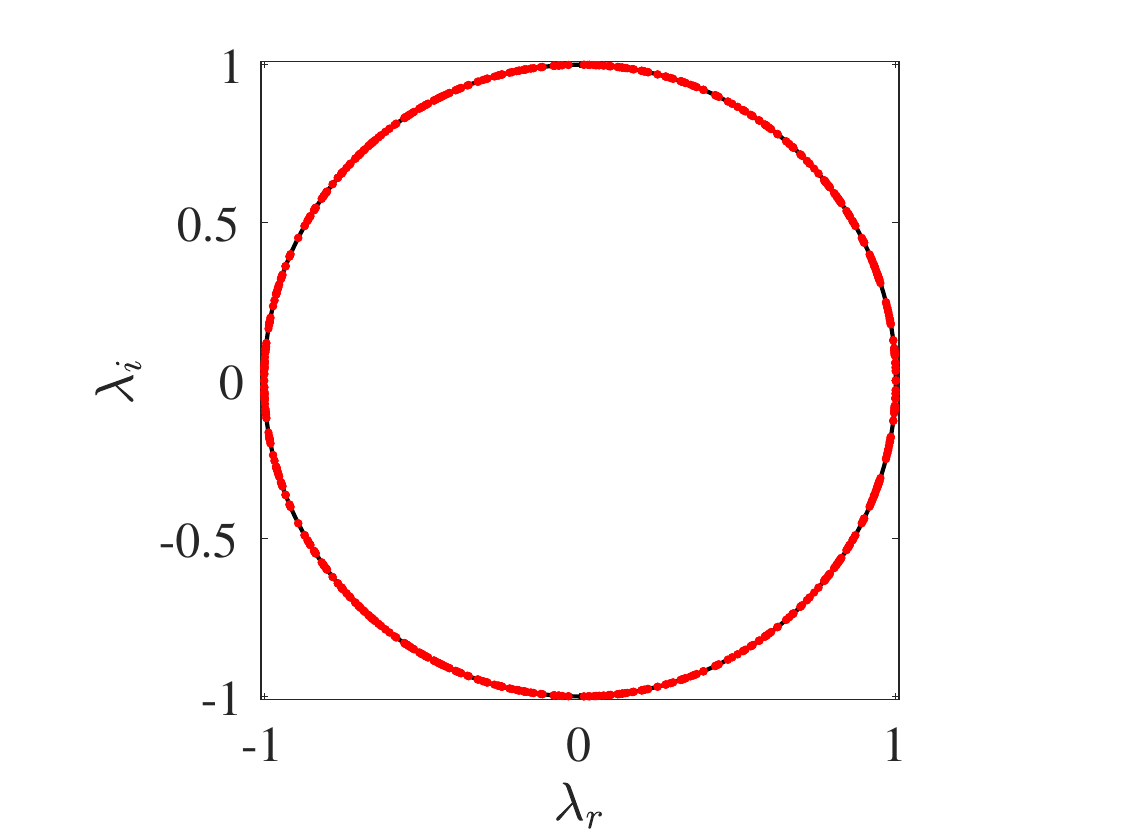}
\put(23.2,62){{\small (b)}}
\end{overpic}
\begin{overpic}[height=.177\textheight, angle =0]{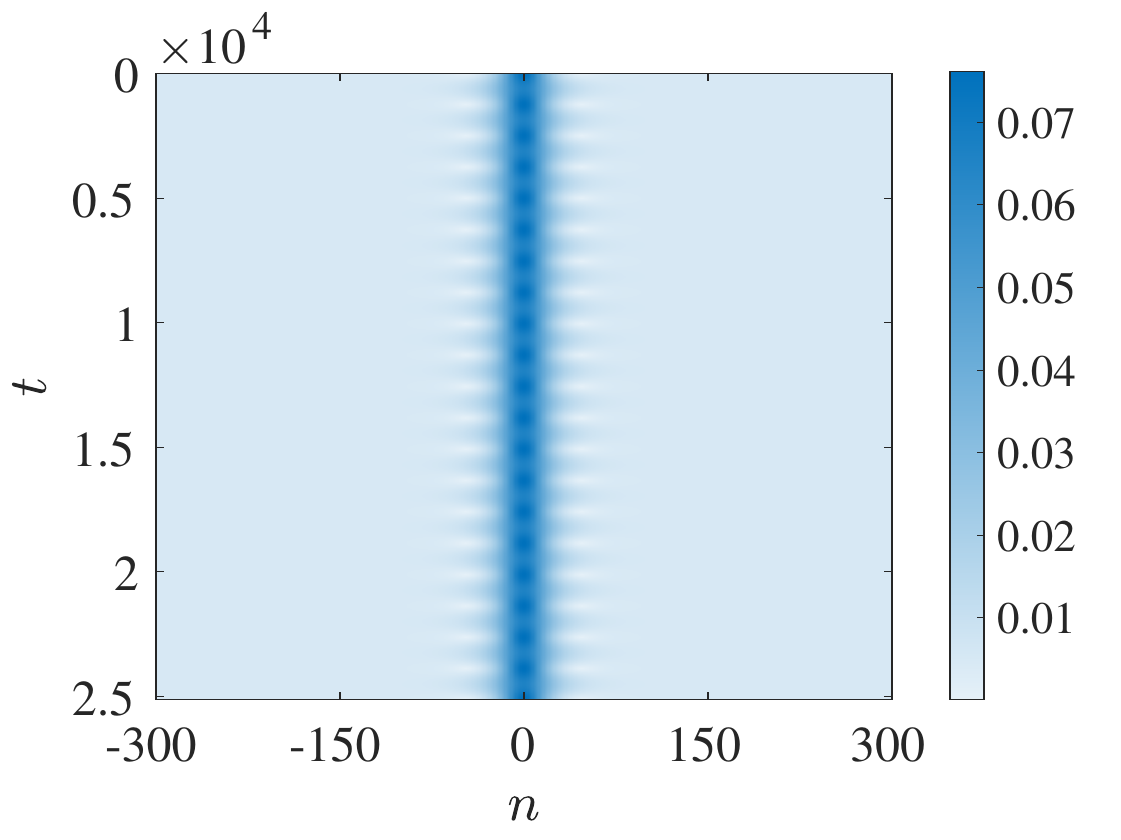}
\put(17,61){{\small (c)}}
\end{overpic}\\
\vskip 0.05cm
\hspace*{-1.7cm}
\begin{overpic}[height=.17\textheight, angle =0]{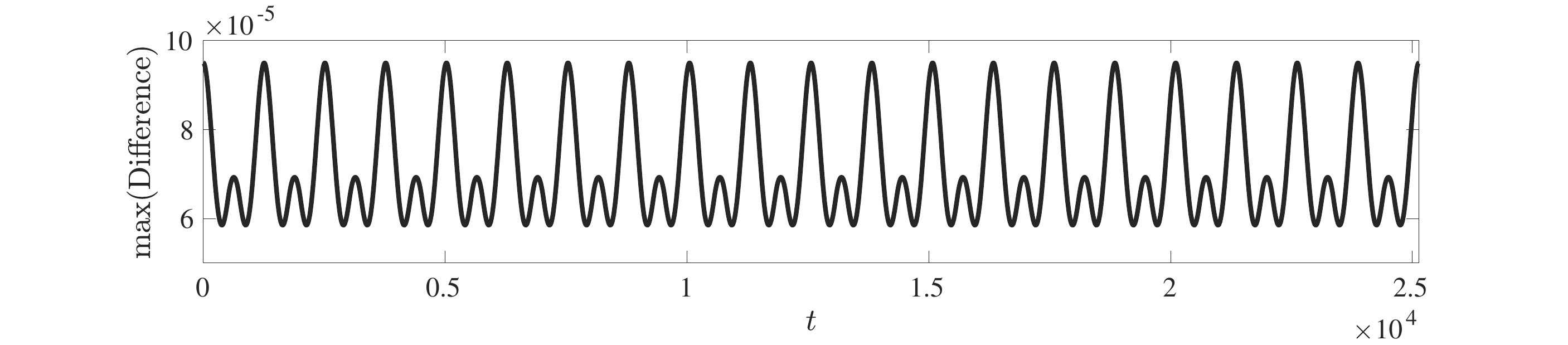}
\put(13.5,17.5){{\small (d)}}
\end{overpic}
\caption{
(Color online)
Summary of results for the numerically exact, KM solution to the DNLS with $q=\omega_{b}=0.005$.%
~The converged KM profile is presented in panel (a) where the spatial distribution of its
amplitude is shown.~The respective Floquet stability analysis results are presented in panel (b)
where a real yet very weakly unstable mode exists with magnitude $\approx 1.000008601445043$ (data
now shown).~The spatio-temporal evolution of the profile of panel (a) is shown in panel (c) showcasing
its robustness over $20$ periods.~Finally, the maximum difference of the amplitudes at each time
instant between the exact KM solution of Eq.~\eqref{km_exact} and numerically evolved KM of the DNLS
equation is shown in panel (d). % (in a semi-logarithmic scale) 
\label{prelim_numer_2}}
\end{figure}

Upon using the exact KM breather of Eq.~\eqref{km_exact} as an initial guess for Newton's
method, our nonlinear solver converged to the profile shown in Fig.~\ref{prelim_numer_2}(a).
% where the spatial distribution of its amplitude is shown. %<- repeats content in caption
~We performed a Floquet stability
analysis of this solution following the setup of the variational equations for the associated
monodromy matrix as discussed in~\cite{sullivan}.~The respective results are shown in Fig.~\ref{prelim_numer_2}(b)
where the red dots correspond to the Floquet multipliers $\lambda=\lambda_{r}+\ii\lambda_{i}$.%
~We find that all the multipliers lie on the unit circle (depicted with a solid black line)
except for a real eigenvalue with a tiny positive real part (data not shown) associated
with a real instability. %<- want to add?
~Per the given lattice of $N=600$ sites and temporal discretization
with $83$ Fourier modes, this instability appears to be rather weak as its magnitude is $\approx 1.0000086$.%
~We check this finding in Fig.~\ref{prelim_numer_2}(c) which showcases the spatio-temporal evolution
of the amplitude $|\psi_{n}(t)|$ of the numerically exact KM breather of Fig.~\ref{prelim_numer_2}(a)
over $20$ periods.~The solution itself is quite robust, maintaining its general characteristics over
the time interval of integration of the DNLS we considered.~We perform a comparison between the
evolution of the exact KM breather of Eq.~\eqref{km_exact} and the numerically obtained one for the DNLS
model in Fig.~\ref{prelim_numer_2}(d).~In particular, we monitor the maximum difference (in
its absolute value) of the amplitude of the exact KM solution and numerically exact KM one to the
DNLS at each time instant.%, and in a semi-logarithmic scale.
~It can be discerned from the panel that
despite being time-periodic solutions to different models themselves, i.e., AL and DNLS, they are
quite proximal to one another (notice the order of the infinity norm of the difference, i.e., $\sim 10^{-5}$).

\begin{figure}[pt]
\begin{center}
\includegraphics[height=.177\textheight, angle =0]{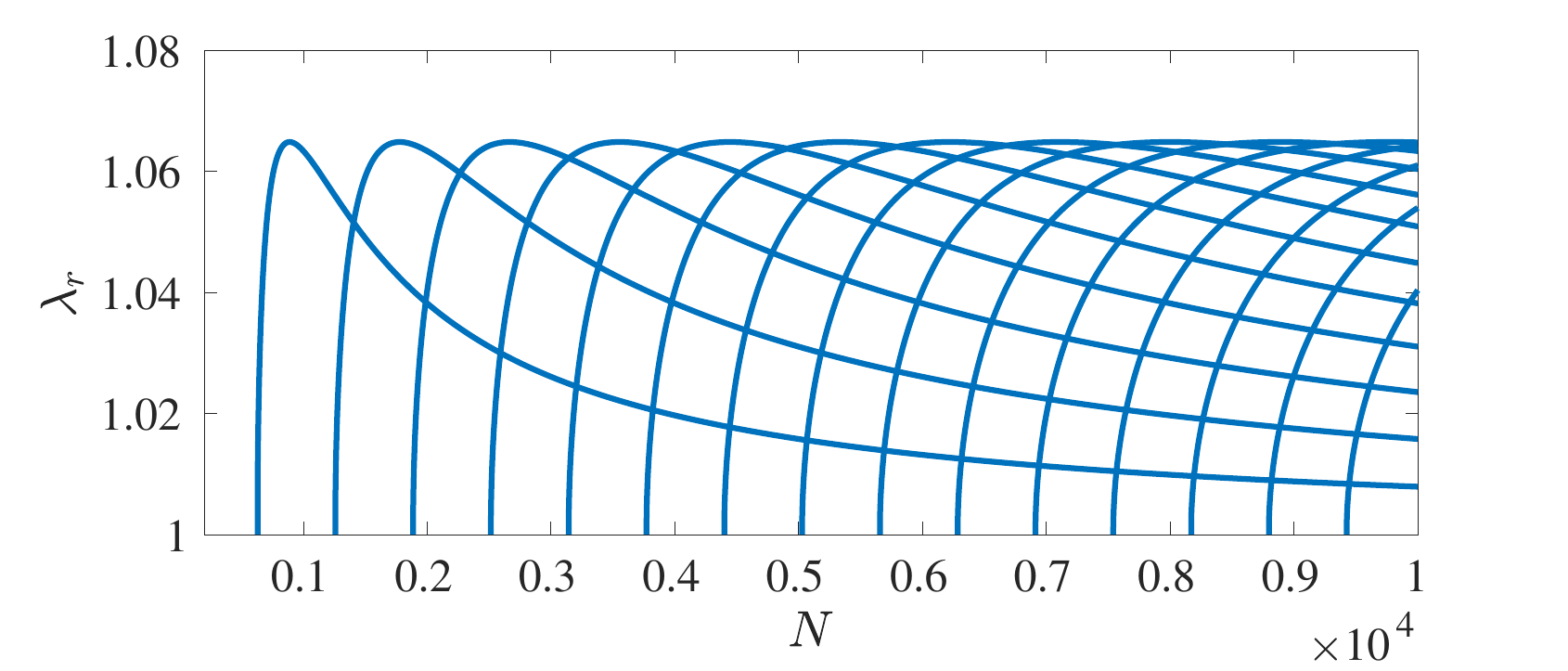}
\end{center}
\caption{
(Color online) MI analysis of the CW solution for $q=\omega=0.005$.~In
particular, real unstable Floquet multipliers, i.e., $\lambda_{r}>1$
(and $\lambda_{i}=0$) are probed as a function of the number of lattice
sites $N$.~The first unstable mode of CW solutions appears when
$N>628$, and more and more unstable modes emerge with increasing $N$.
\label{prelim_numer_3}
}
\end{figure}

% \subsection{A comment o}\label{sec_2_1}
% % ======================================================================
Finally, we would like to comment on an important observation of our spectral stability analysis
results.~It is known from~\cite{kivpey}, that plane wave solutions (CW) to the focusing DNLS model
are modulationally unstable (a stability calculation of CW solutions to the Salerno model that contains
the AL is presented in~\cite{sullivan} too).~The KM breathers (alongside with P and AB ones) are
mounted atop a finite yet modulationally unstable background (per the focusing DNLS and AL) where
localized solutions may perturb the CW Floquet spectrum.~In light of the spectral stability analysis
results of Fig.~\ref{prelim_numer_2}(b) and the discretization used in the present calculation,
remarkably, no unstable modes are observed other than the weakly unstable one mentioned previously.%
~We also note that the time-translation mode $1+0\ii$ is accurately resolved in our computations.%

We investigated the absence of instabilities of the KM breather of Fig.~\ref{prelim_numer_2}(a) by
performing a modulational instability (MI) analysis of the CW with $q=\omega_{b}=0.005$ parametrically as
a function of the number of sites $N$.~Our respective results are shown in Fig.~\ref{prelim_numer_3} where
we probe the real unstable modes $\lambda_{r}$ as a function of $N$.~In conjuction with the spectrum of
Fig.~\ref{prelim_numer_2}(b) with $N=600$, the MI analysis for the same $N$, surprisingly does not predict
the presence of CW unstable modes.~In fact, this is true for all $N\lesssim 628$.~However, for a value of
$N$ just above $N=628$, we observe the emergence of the first unstable mode, and in general, the number of
unstable modes starts getting increased with $N$ [cf.~Fig.~\ref{prelim_numer_3}], i.e., we obtain a \textit{band of
unstable modes} as we approach the infinite lattice.~This finding in turn suggests that for solutions
sitting atop a finite yet \textit{small} background, the computation of their Floquet spectrum requires the
use of a very large number of sites for resolving it well.
% ~{Equipped with this observation}, %This way, 
% and despite the absence of real unstable
% modes in the Floquet %spectra of KM breathers in 
% {spectrum of the KM breather of} Fig.~\ref{prelim_numer_2}(b) but also in the spectra presented
% next, the KM solutions that we identify in this work are expected to be (at least) MI unstable {as $N$ increases}.
~Despite the absence of a band of real unstable
modes in the Floquet %spectra of KM breathers in 
{spectrum of the KM breather of} Fig.~\ref{prelim_numer_2}(b) and in the spectra presented
next, this observation indicates that the KM solutions that we identify in this work are expected to be  MI unstable {as $N$ increases}.

Having presented our preliminary numerical results
to motivate the notion of proximity between the AL
and DNLS models for small enough data, we 
% rigorously 
 analytically %*** check
explore this proximity
%, i.e.,
%distance between the solutions to the AL and DNLS systems 
in the next section.

%%%%
\section{Theoretical justification of the proximal KM dynamics by DNLS}
\subsection{The case of the infinite lattice}
% For our subsequent theoretical analysis, we will consider the Salerno model of Eq.~\eqref{salerno0}
% where we will write separately the AL and DNLS models respectively.
For our subsequent theoretical analysis, we will introduce notation to distinguish  the AL and DNLS limits of the Salerno model of Eq.~\eqref{salerno0}.
~Recall that the AL model corresponds
to the $g=1$ case:
\begin{equation}
	\ii\dot{\Psi}_{n}+C(\Delta_d\Psi)_n%
	+\left(\Psi_{n+1}+\Psi_{n-1}\right)|\Psi_{n}|^{2}=0,
	\label{cAL}
\end{equation}
whereas the DNLS model to the $g=0$ case:
\begin{equation}
	\ii\dot{\Phi}_{n}+C(\Delta_d\Phi)_n%
	+2|\Phi_{n}|^{2}\Phi_n=0,
	\label{cDNLS}
\end{equation}
that is, $\Psi$ and $\Phi$ will stand for the solutions to the AL and DNLS
models, respectively.~Motivated by \cite{JDE1}, the systems will be supplemented
with general non-zero boundary conditions of the form
\begin{equation}\label{nbc1}
	\lim_{|n|\rightarrow\infty} \Psi_n(t) = \lim_{|n|\rightarrow\infty}e^{2i q^2t}q_n, 
	\quad 
	\lim_{|n|\rightarrow\infty} \Phi_n(t) = \lim_{|n|\rightarrow\infty} e^{2iq^2t} q_n, \quad t\geq 0,
\end{equation}
where 
\begin{equation}
	\label{nbc2}
q_n = \left\{
\begin{array}{ll}
	q_-, &n\leq 0,
	\\
	q_+, &n>0,
\end{array}
\right.
\end{equation}
and $q_{\pm}$ are complex constants with $|q_{\pm}|=q$.
The boundary conditions of Eqs.~\eqref{nbc1}-\eqref{nbc2} can be transformed to be
time independent via the change of variables
\begin{equation}\label{nbc3}
	\Psi_n(t) = e^{2iq^2t} \psi_n(t), 
	\quad 
	\Phi_n(t) = e^{2iq^2 t} \phi_n(t).
\end{equation}
%
% With the change of variables of Eq.~\eqref{nbc3}, then Eqs.~\eqref{cAL} and \eqref{cDNLS}
% are written in the form
Using the change of variables in Eq.~\eqref{nbc3}, we can rewrite Eqs.~\eqref{cAL} and \eqref{cDNLS} 
in the form
\begin{align}
	\label{ncAL}
	&i \dot{\psi}_n + C (\Delta_d \psi)_n- 2q^2 \psi_n+ |\psi_n|^2\left(\psi_{n+1}+\psi_{n-1}\right)=0,
	\quad
	\\
	&i\dot{\phi}_n + C (\Delta_d \phi)_n+ 2 \left[|\phi_n|^2 - q^2\right] \phi_n = 0.
	\label{ncDNLS}
\end{align}
%
% Note that when considering the initial conditions, these 
Note that the initial conditions
remain unchanged under the
change of variables of Eq.~\eqref{nbc3}, namely, $\psi_n(0) = \Psi_n(0)$ and
$\phi_n(0) = \Phi_{n,0}$. The  boundary conditions become
\begin{equation}\label{nbc4}
	\lim_{n \to \pm \infty} \psi_n(t) = \lim_{n\to \pm \infty} \phi_n(t)
	=
	q_\pm,
\end{equation}
so that $\displaystyle{\lim_{|n|\rightarrow\infty} |\psi_n(t)| = \lim_{|n|\rightarrow\infty} |\Phi_n(t)| = q>0}$.
Next, we perform a second change of variables, 
\begin{equation}\label{nbc5}
	\psi_n(t) = U_n(t)+q_n,
	\quad
	\phi_n(t) = V_n (t)+q_n,
\end{equation}
and Eqs.~\eqref{ncAL}-\eqref{ncDNLS} become
\begin{align}
	\label{dbcAL}
	i\dot{U}_n+C (\Delta_d U)_n&+C (\Delta_d q)_n -2 q^2(U_n+q_n)\nonumber\\
 &+ |U_n+q_n|^2\left(U_{n+1}+U_{n-1}+q_{n+1}+q_{n-1}\right)=0,
	\\
	\label{dbcNLS}
	i \dot{V}_n+ C (\Delta_d V)_n &+C (\Delta_d q)_n +2 \left[|V_n+q_n|^2 - q^2\right] (V_n+q_n)=0.
\end{align}
The initial conditions for the equations \eqref{dbcAL}-\eqref{dbcNLS}, become
\begin{equation}\label{mod-ic}
	U_n(0) = \psi_n(0) - q_n,
	\quad
	V_n(0) = \phi_n(0) - q_n,
\end{equation}
and the boundary conditions are \textit{zero}   at infinity, i.e.,
\begin{equation}\label{vbcn}
	\lim_{n\rightarrow\pm\infty}U_n(t)
	=
	\lim_{n\rightarrow\pm \infty}V_n(t) = 0,\quad t\geq 0.
\end{equation}

Along the lines of \cite{Fotop2024} (see also \cite{JDE1} for the continuous counterpart), we can prove that, for any initial conditions $U(0),V(0)\in \ell^2$, there exist $T_{AL}^*(U(0)), T_{DNLS}^*(V(0))>0$ such that the above Cauchy problems for the modified AL and DNLS equations \eqref{dbcAL} and \eqref{dbcNLS} have unique solutions   $U\in C^1([0,T_{AL}^*(U(0))],\ell^2)$   and $V\in C^1([0,T_{DNLS}^*(V(0))],\ell^2)$.
Starting from this local existence result and following the methods of \cite{JDE0,JDE2,JDE1}, we can further prove that these solutions stay close to one another in the following sense:

\begin{theorem}
\label{Clos}
Consider the Cauchy problems for the modified AL and DNLS equations \eqref{dbcAL} and \eqref{dbcNLS} when supplemented with the initial conditions \eqref{mod-ic} and the vanishing boundary conditions at infinity \eqref{vbcn}. Let any $0<\varepsilon<1$ and assume that the initial conditions $U(0)$ and $V(0)$ have $\ell^2$-norms of $\mathcal{O}(\varepsilon)$, the $\ell^2$-distance $||U(0)-V(0)||_{\ell^2}$ between them is of $\mathcal{O}(\varepsilon^3)$, and that the background amplitude $q$ is of  $\mathcal{O}(\varepsilon)$, namely there exist constants $C_i,\;i=1,...,4$
\begin{eqnarray}
\label{sd}
||U(0)||_{\ell^2}\leq C_1\varepsilon,\;\;||V(0)||_{\ell^2}\leq C_2\varepsilon,\;\;||U(0)-V(0)||_{\ell^2}\leq C_3\varepsilon^3,\;\;\mbox{and}\;\;q\leq C_4\varepsilon.
\end{eqnarray}
Then, there exists $T_c>0$ and a positive constant $C>0$ such that the $\ell^2$-distance between the solutions satisfies simultaneously the upper bounds
\begin{equation}
\label{closinf}
\begin{aligned}
 &||U(t)-V(t)||_{\ell^2}\leq C\varepsilon^3t,
 \\
 &||U(t)-V(t)||_{\ell^2}\leq \hat{C}\varepsilon,
  \end{aligned}
 \quad\mbox{for all}\;\; 0<t\leq T_c.
\end{equation}
\end{theorem} 
Theorem \ref{Clos} is proved in  \ref{app}.
The time $T_c$ in the above proximity result is obtained as follows. Since $U\in C^1([0,T_{AL}^*],\ell^2)$, i.e., it is continuously differentiable with respect to time, the assumption \eqref{sd} implies that there exists a time $\hat{T}_{AL} \in (0, T_{AL}^*]$ such that the size of $U(t)$ is also of  $\mathcal{O}(\varepsilon)$, that is
\begin{eqnarray}
\label{cruc1a}  
||U(t)||_{\ell^2}\leq C_5\varepsilon,\;\;\mbox{for all}\;\;t\in [0, \hat{T}_{AL}]
\end{eqnarray}
for some constant $C_5>0$. Similarly, there is time $\hat{T}_{DNLS} \in (0,T_{DNLS}^*]$ such that 
\begin{eqnarray}
\label{cruc1b}  
||V(t)||_{\ell^2}\leq C_6\varepsilon,\;\;\mbox{for all}\;\;t\in [0, \hat{T}_{DNLS}],
\end{eqnarray}
for some constant $C_6>0$. Then, the proximity time $T_c$ is defined as $T_c=\min\big\{\hat{T}_{AL},\hat{T}_{DNLS}\big\}$.

\subsection{The case of periodic boundary conditions}

As described in Section \ref{sec_2}, in the numerical simulations we supplement the lattice \eqref{salerno} with periodic boundary conditions. This scenario bears
some significant differences and is more tractable than the case of the infinite lattice discussed above, due to the conservation laws of the systems.
The phase spaces for the periodic lattice are  the spaces of periodic sequences with  period $N$,  denoted by
\begin{equation*}
	{\ell}^p_{per}:=\Big\{U=(U_n)_{n\in\mathbb{Z}}\in\mathbb{R}:\  U_n=U_{n+N},\ \|U\|_{\ell^p_{\mathrm{per}}}:=\Big(h\sum_{n=0}^{N-1}|U_n|^p\Big)^{\frac{1}{p}}<\infty\Big\}, \  1\leq p\leq\infty,
\end{equation*}	
where $h$ denotes the lattice spacing. For simplicity, we set $h=1$ (as in the numerical study which corresponds to the choice $C=\frac{1}{h^2}=1$ made above). 
In particular, we have the following analogue of Theorem \ref{Clos}, which can be proved in exactly the same way as the results given in \cite{JDE0,JDE2}.

\begin{theorem}
\label{ClosPer}
Consider the Cauchy problems for equations \eqref{ncAL} and \eqref{ncDNLS} when supplemented with periodic boundary conditions. Let any $0<\varepsilon<1$ and assume that the initial conditions $\psi(0)$ and $\phi(0)$ have $\ell^2_{per}$-norms of $\mathcal{O}(\varepsilon)$, the $\ell^2_{per}$-distance $||\psi(0)-\phi(0)||_{\ell^2_{per}}$ between them is of $\mathcal{O}(\varepsilon^3)$, and the background amplitude $q$ is of  $\mathcal{O}(\varepsilon)$, i.e. there exist positive constants $C_i$, $i=1,...,4$ such that 
\begin{eqnarray}
\label{sdper}
||\psi(0)||_{\ell^2_{per}}\leq C_{1}\varepsilon,\;\;||\phi(0)||_{\ell^2_{per}}\leq C_{2}\varepsilon,\;\;||\psi(0)-\phi(0)||_{\ell^2_{per}}\leq C_3\varepsilon^3,\;\;\mbox{and}\;\;q\leq C_4\varepsilon.
\end{eqnarray}
Then, for arbitrary $0<T<\infty$, there exists a constant $\tilde{C}>0$ such that the $\ell^2_{per}$-distance between the solutions satisfies the estimate
\begin{eqnarray}
\label{BDPER}
 ||\psi(t)-\phi(t)||_{\ell^2_{per}}\leq \tilde{C}\varepsilon^3t,\;\;\mbox{for all}\;\; 0<t\leq T.  
\end{eqnarray}
\end{theorem}

The main difference between Theorems \ref{Clos} and \ref{ClosPer} is that in the latter case the conservation laws ensure the global existence of solutions, thus implying the validity of the proximity result \eqref{BDPER} for arbitrary times. %***

\subsection{Connection to the numerical findings of Section \ref{sec_2}}

Theorem \ref{Clos} (for the infinite lattice with nonzero boundary conditions) and Theorem \ref{ClosPer} (for the finite lattice with periodic boundary conditions) justify theoretically that the distance between the solutions of the AL and DNLS equations grows linearly at a rate which is at most of $\mathcal{O}(\varepsilon^3)$ when the initial and background data are of $\mathcal{O}(\varepsilon)$.

\begin{figure}[pt!]
\centering
\includegraphics[scale=0.8]{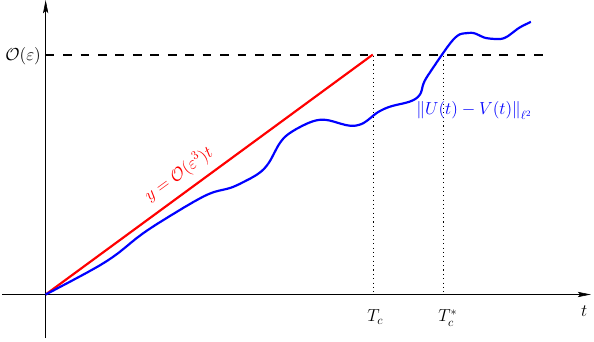}
\includegraphics[scale=0.8]{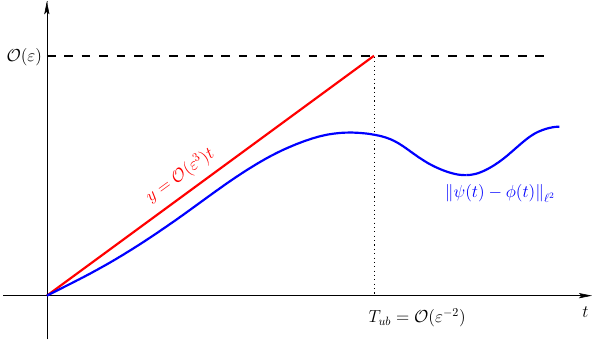}
\caption{The behavior of the distance according to the theoretical upper bounds of Theorems \ref{Clos}-\ref{ClosPer}. \textit{Left:} The case of nonzero boundary conditions. Here, the distance of solutions may increase beyond $\mathcal O(\varepsilon)$ at finite time. \textit{Right:} The case of periodic boundary conditions. Here, according to the uniform bound of inequality~\eqref{tri1}, the distance of solutions is at most $\mathcal O(\varepsilon)$ for all times. It should be noted that, in certain scenarios,  the distance curve (in blue) may be much closer to the $t$-axis than what is depicted above, at least for an initial time interval --- e.g. see the corresponding graphs of Fig.~\ref{numerics_proximity}.}
\label{proximity_fig}
\end{figure}

In the case of Theorem \ref{Clos}, according to the well-posedness results of \cite{JDE1,hkmm2025},  initial data and background of $\mathcal O(\varepsilon)$ guarantee that the $\ell^2$ norm of the solutions remains of $\mathcal O(\varepsilon)$ over a lifespan of  $\mathcal O(\varepsilon^{-2})\simeq T_c$. However, beyond that lifespan, the solutions may not remain of  $\mathcal O(\varepsilon)$ and, therefore, there may exist a time $T_c^*$ such that the distance of solutions may escape from the trapezoidal region delimited by the two upper bounds in \eqref{closinf}. 
This scenario is portrayed by the left diagram of Fig.~\ref{proximity_fig}. 

On the other hand, in the case of Theorem \ref{ClosPer} for the periodic problem, the $\mathcal O(\varepsilon^3)$ growth rate of the difference of solutions is \textit{transient}. 
Indeed, recalling the conserved quantities \eqref{eq:consv_AL} and \eqref{eq:consv_DNLS}, it was shown in \cite{JDE0} that
\begin{eqnarray}
		||\psi(t)||_{\ell^2_{per}}^2 \le \exp(P_{\mathrm{AL}}(0))-1\lesssim\varepsilon^2,\quad \text{for all } t\ge 0,\label{Bound1}
\end{eqnarray}
and, furthermore, due to the conservation of $P_{\mathrm{DNLS}}$,
\begin{eqnarray}
\label{consDNLS}
P_{\mathrm{DNLS}}(t)=||\phi(t)||^2_{\ell^2_{per}}=||\phi(0)||^2_{\ell^2_{per}}=P_{\mathrm{DNLS}}(0)\lesssim \varepsilon^2,\quad \text{for all } t\geq 0.
\end{eqnarray}
Hence, by the triangle inequality, 
\begin{eqnarray}
 \label{tri1}
||\psi(t)-\phi(t)||_{\ell^2_{per}}\leq
||\psi(t)||_{\ell^2_{per}}+||\phi(t)||_{\ell^2_{per}}\leq \hat{C}\varepsilon,\;\;\mbox{for all $t\geq 0$},
\end{eqnarray}
where the constant $\hat{C}$ is independent of $t$. Consequently, the linear growth upper bound in the proximity estimate \eqref{BDPER} is relevant only for times such that
% can never exceed the uniform in time upper bound \eqref{tri1}. Therefore, it must be that
$\tilde{C}\varepsilon^3t\leq \hat{C}\varepsilon$, i.e. 
\begin{eqnarray}
  \label{tri2}
  t\leq \frac{\hat{C}}{\tilde{C}}\frac{1}{\varepsilon^2}=:T_{ub}.
\end{eqnarray}
That is, $T_{ub}$ defines the upper bound for the times where the linear growth of the distance between the solutions holds. For $t>T_{ub}$, the distance between the solutions should satisfy the uniform upper bound \eqref{tri1}. This situation is illustrated in the right diagram of Fig.~\ref{proximity_fig}.

Under the $\mathcal{O}(\varepsilon)$ smallness conditions  on the initial and background data \eqref{sd} and \eqref{sdper},   Theorem  \ref{Clos}  for the infinite lattice supplemented with the nonzero boundary conditions \eqref{nbc1}-\eqref{nbc2} and Theorem \ref{ClosPer} for the periodic lattice  justify  that the DNLS lattice admits solutions of $\mathcal{O}(\varepsilon)$ with the following properties:
\begin{enumerate}
\item They diverge at most linearly in time
from the analytical solutions of the AL lattice in terms of  the $\ell^2$-metric, with a linear growth rate at most of $\mathcal{O}(\varepsilon^3)$, for finite times in $[0, T_c]$ in the case of the infinite lattice (left panel of Fig.~\ref{proximity_fig}) or satisfying at most the upper bound $T_{ub}$ given by \eqref{tri2} in the periodic case (right panel of Fig.  \ref{proximity_fig}) .
\item For all $t\geq 0$, the solutions remain close --- at most of $\mathcal{O}(\varepsilon)$ with respect to the $\ell^2$-metric --- for all $t\in [0, T_c]$ in the case of the infinite lattice and for all $t>0$ in the periodic case.
\item For the distance measured in all $\ell^p$-metrics with $p\geq 2$, due to the embedding
\begin{equation}
	\label{lp1}
	\ell^q\subset\ell^p,\quad \|u\|_{\ell^p}\leq \|u\|_{\ell^q}, \quad 1\leq q\leq p\leq\infty,
\end{equation}	
it is reasonable to expect an even smaller rate of proximity than the one for the $\ell^2$-metric for the times described above. For example, when $p=\infty$, which is relevant for a comparison of the amplitudes of solutions, it is theoretically justified to expect an improved rate of proximity.
\end{enumerate}

In summary, comparing the results of Theorems \ref{Clos} and \ref{ClosPer}, in the case of the infinite lattice there is a \textit{possibility} that, at some finite time $T_c^* \geq T_c \simeq \mathcal O(\varepsilon^{-2})$, the distance of solutions will escape from the trapezoidal region capped  by the $\mathcal O(\varepsilon)$ dashed horizontal line (see Fig.~\ref{proximity_fig}).  This is due to the fact that, in the case of nonzero boundary conditions, it is not known whether the solutions are uniformly bounded. On the other hand, in the  case of periodic boundary conditions, the solutions individually remain bounded by the initial data of size $\mathcal O(\varepsilon)$ due to the conservation of $P_{\text{DNLS}}$ and $P_{\text{AL}}$; therefore, by the triangle inequality, the distance of solutions is guaranteed to remain within the trapezoidal region below the $\mathcal O(\varepsilon)$ dashed horizontal line at all times (see Fig.~\ref{proximity_fig}).

It is remarkable that the numerical findings of Section \ref{sec_2} for the case of the KM solutions showcase that the growth rate of the distance is significantly lower than the ones associated with the theoretical bounds, and the order of proximity is much higher than what is predicted theoretically.~This is in accordance with several numerical findings for both discrete \cite{JDE0,JDE2} and continuous \cite{JDE1} setups, which illustrate that the rate of divergence and proximity may depend on the particular analytical solution of the integrable system considered. In many cases (bright solitons \cite{JDE0}, fast soliton collisions \cite{JDE1}), this rate is of $\mathcal{O}(\varepsilon^p)$ with $p>3$.%
~Indicatively, and in line with the numerical study of Fig.~\ref{prelim_numer}, 
the rate of divergence of the solutions was found to be in the neighborhood of $\varepsilon^5$.
%have that $\|\psi(0)\|_{\ell^{1}}\equiv\|\phi(0)\|_{\ell^{1}}\approx 0.076$, and \eqref{sdper}}
%In the  case of the numerical study of Fig.~\ref{prelim_numer}, the linear growth rate was found to be of $\mathcal{O}(\varepsilon^5)$.~In particular, the third property discussed above justifies the small order of the difference of solutions in the 
%$\ell^{\infty}$-norm (which was found to be of $\mathcal{O}(10^{-5})$ as reported at the end of Section \ref{sec_2}).
%PGK: yes *but* 10^{-5} should be compared with
% what theoretically?? I.e., what is \epsilon in this
% case. Presumably \epsilon should be sth like 0.075?
% Notice that in this case (as in others we have
% considered practically) C3=0 effectively...
% If you agree, I would say here that the difference
% was found to be in the neighborhood of \epsilon^4
% or similar if I understand correctly...
% \epsilon=0.075 (max of solution).
% The difference between solutions is bounded by \varepsilon^{3} for all times.
\begin{figure}[pt!]
\centering
\begin{overpic}[height=.16\textheight, angle =0]{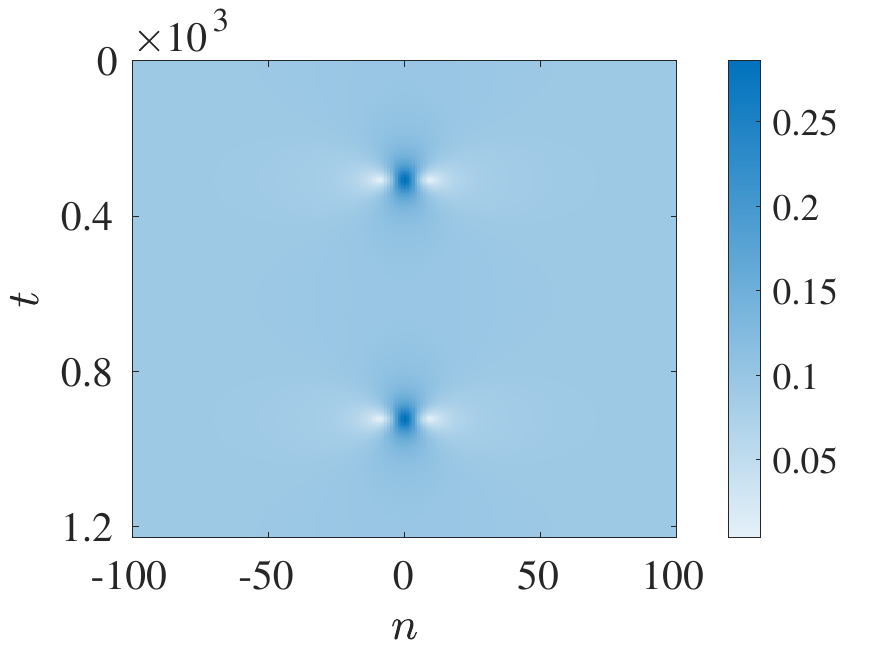}
\put(18,60){{\small (a)}}
\end{overpic}
\begin{overpic}[height=.16\textheight, angle =0]{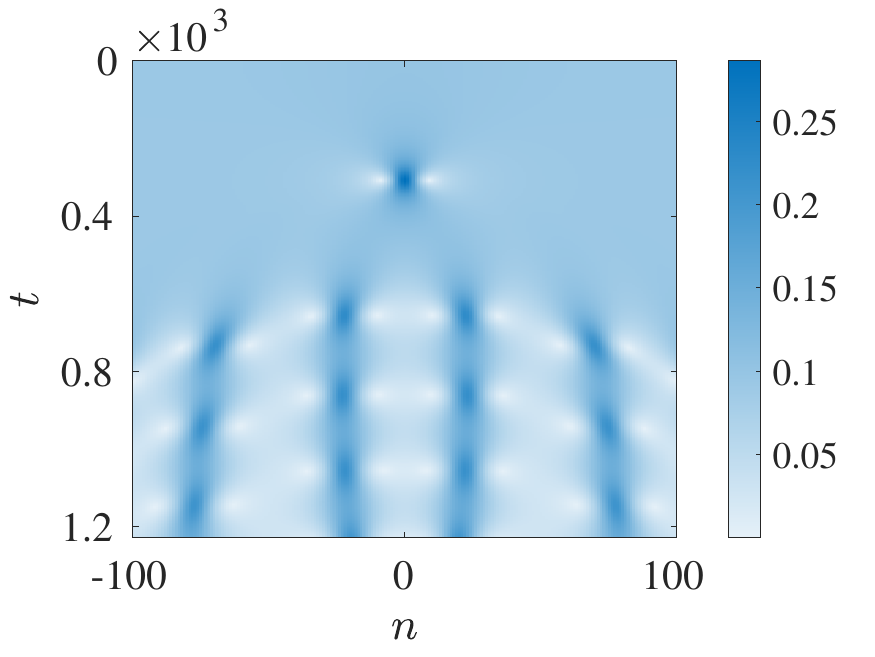}
\put(18,60){{\small (b)}}
\end{overpic}
\begin{overpic}[height=.16\textheight, angle =0]{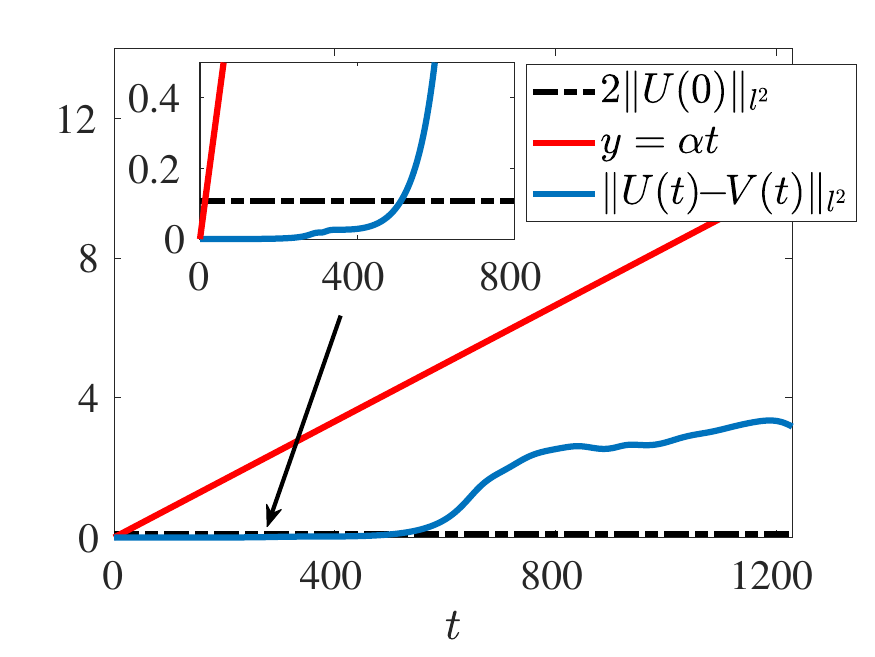}
\put(80,35){{\small (c)}}
\end{overpic}
\begin{overpic}[height=.16\textheight, angle =0]{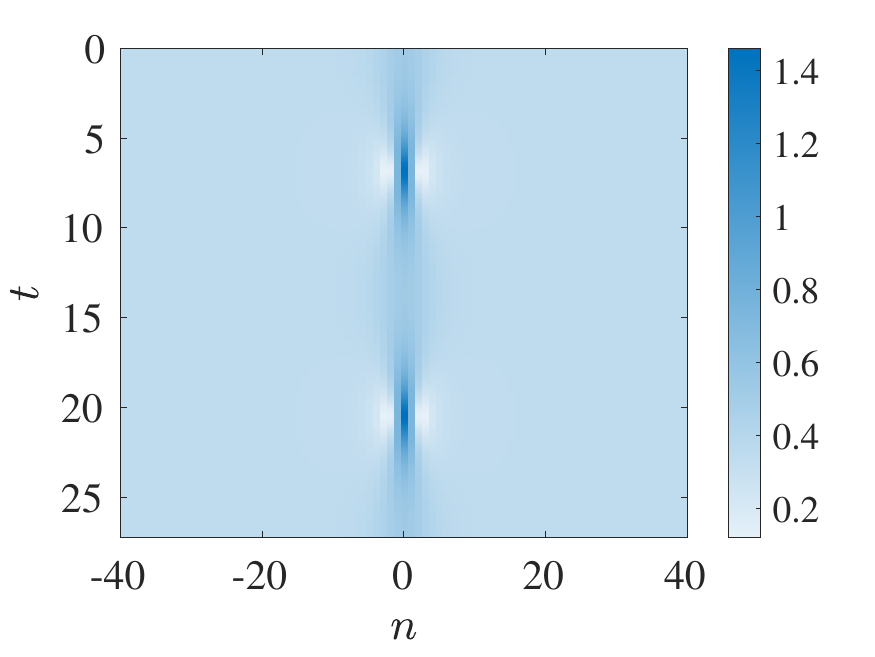}
\put(18,60){{\small (d)}}
\end{overpic}
\begin{overpic}[height=.16\textheight, angle =0]{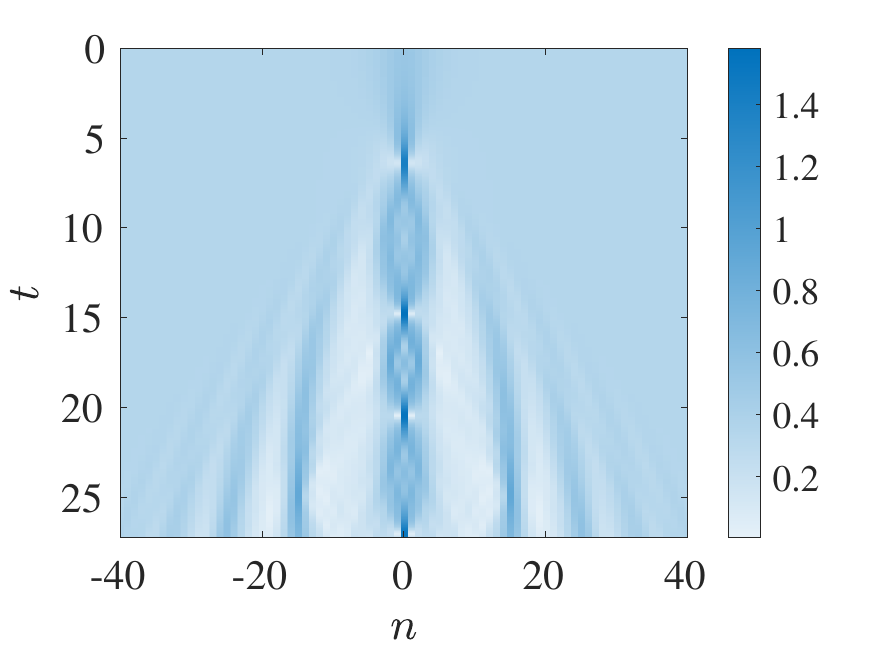}
\put(18,60){{\small (e)}}
\end{overpic}
\begin{overpic}[height=.16\textheight, angle =0]{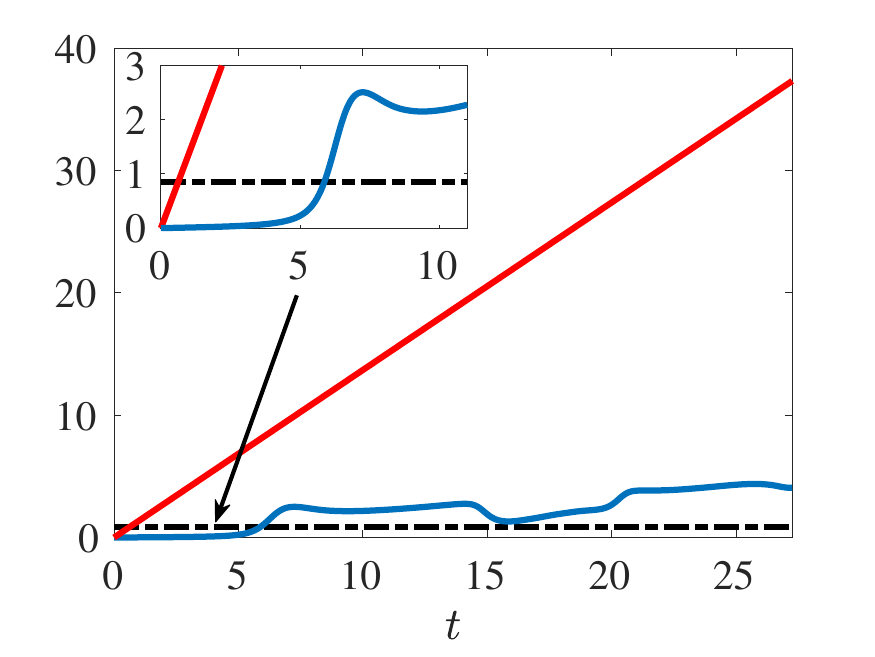}
\put(80,35){{\small (f)}}
\end{overpic}
\caption{
%
%   Top row:
% =============================================
%       k_{2} = -0.05330541914895709 (=r)
%           q = 0.09170814099744679
%  \omega_{b} = 0.010228994294878715
%       T_{b} = 614.2524989309406
%   \|U(0)\|_{l^{2}} = 0.053923830354692
%  \alpha = 3*epsilon^3+9*q*epsilon^2+12*q^2*epsilon = 0.008312640151497
%
% KM appears the first time for the DNLS when t\approx 308. The distance
% crosses the 2\|U(0)\|_{l^{2}} at t\approx 512
%
%
%   Bottom row:
% =============================================
%       k_{2} = -0.5 (=r)
%           q = 0.3400008782159792
%  \omega_{b} = 0.4612603846816791
%       T_{b} = 13.621775283207297
%   \|U(0)\|_{l^{2}} = 0.424655262285145
%  \alpha = 3*epsilon^3+9*q*epsilon^2+12*q^2*epsilon = 1.370639364437760
%
% KM appears the first time for the DNLS when t\approx 6.36. The distance
% crosses the 2\|U(0)\|_{l^{2}} at t\approx 5.86.
%
Summary of numerical results for the AL and DNLS models [cf.~Eqs.~\eqref{ncAL} and~\eqref{ncDNLS}]
in conjuction with our theoretical analysis on their proximity.~We evolve KM initial data
with $(q,\omega_{b},\varepsilon)\simeq(0.09,0.01,0.054)$ and $(q,\omega_{b},\varepsilon)\simeq(0.34,0.46,0.425)$
in the top and bottom rows, respectively, over two periods.~Panels (a) and (d), and (b)
and (e) depict the spatio-temporal evolution of the amplitudes $|\psi_{n}|$ and $|\phi_{n}|$,
respectively.~The panels (c) and (f) summarize the comparisons made between the theoretical
estimates on the proximity of the models with the numerically obtained distance (see, the legend
and insets therein) of the solutions after subtracting the background $q$ (see, text for details).
\label{numerics_proximity}}
\end{figure}

We conclude our theoretical analysis by showcasing two examples comparing the theoretical
estimates with the numerical computations of Fig.~\ref{numerics_proximity}.
Note that, for computational purposes, we may consider the assumptions \eqref{sd} with constants $C_i = 1$, $i=1,...,4$. Then, we may consider the relevant times 
$\hat{T}_{AL},~\hat{T}_{DNLS}$ such that the bounds \eqref{cruc1a} and \eqref{cruc1b} hold with the constants $C_5=C_6=1$. 
% In this case, we may derive an explicit form of the closeness estimate \eqref{closinf}, by inserting the estimates \eqref{cruc1a} and \eqref{cruc1b} in \eqref{GAL} and \eqref{GDNLS}, respectively, for the above choice of unit constants $C_i$.
For this choice of unit constants $C_i$, we may derive an explicit form of the closeness estimate \eqref{closinf} by inserting the estimates \eqref{cruc1a} and \eqref{cruc1b} in \eqref{GAL} and \eqref{GDNLS}, respectively.
This way, by using \eqref{estdist1}, we derive the following explicit form of the bound \eqref{closinf} depending on $\varepsilon$ and $q$,
\begin{eqnarray}
   \label{numest}
||U(t)-V(t)||_{\ell^2}\leq \alpha t,\;\;\alpha=3\varepsilon^3+9q\varepsilon^2+12q^2\varepsilon.
\end{eqnarray}

The top and bottom rows
of Fig.~\ref{numerics_proximity} correspond to the cases with $(q,\omega_{b})=(0.09,0.01)$
and $(0.34,0.46)$, respectively.~Same as before, we employ
a lattice of $N=600$ sites, and use the exact KM solution of Eq.~\eqref{km_exact} as an
initial condition for both the AL and DNLS models.~The panels (a) and (d), and (b) and (e)
showcase the spatio-temporal evolution of the amplitude of the KM solution (in the respective cases)
for the AL and DNLS models, respectively, over 2 periods.~In line with Fig.~\ref{proximity_fig},
we summarize our proximity results in panels (c) and (f) of Fig.~\ref{numerics_proximity}
where the solid blue line (see the legend in panel (c)) shows the distance of the solutions
in the $l^{2}$ norm after removing the background $q$. %as a function of time.~
The horizontal
dashed black line corresponds to  $y=2\|U(0)\|_{l^{2}}=2\varepsilon$, whereas the solid red line to $y=\alpha t$, with $\alpha$ given by \eqref{numest}. For panels (a)-(c), $\varepsilon \simeq 0.054$, while for panels (d)-(f) we used $\varepsilon \simeq 0.425$. Recall that, according to the theoretical results of Theorem \ref{Clos} and \ref{ClosPer}, the solutions should remain proximal for minimal 
guaranteed times of $T_{c}\simeq\mathcal O(\varepsilon^{-2})$.~The numerical results of Figs.~\ref{numerics_proximity}(c)
and (f) confirm this fact, since they show that the two models are proximal up to times $\approx 512$ (top row) and $\approx 5.86$ (bottom row), see
the insets therein.%
%$T_{c}^{\ast}\approx 512$ (top row)
%and $T_{c}^{\ast}\approx 5.86$, see the insets therein. 
%PGK: Can you please revisit the definition of
%T_c and examine whether T_c has numerically been
%computed according to its definition. I am not sure
% that it has been as far as I can tell...
% We can't say that this is the T_{c}.
~Furthermore, the comparison of the patterns over the proximity time scales justify accordingly the proximity of the spatiotemporal dynamics manifested by the emergence of the first apparent 
localization event in each of the panel pairs (a)-(b) and (d)-(e), respectively. Beyond this time, the MI effects manifest themselves
in the evolution of the KM for the DNLS model (see, panels (b) and (e)), along the lines of the results of \cite{blmt2018}.~Those effects contribute to the
increase of the distance between the solutions of the AL and DNLS models, although the latter always 
remains bounded by the linear, theoretically 
predicted growth of Figs.~\ref{numerics_proximity}(c)
and (f); see the respective red lines.
It is also interesting to observe that while
the blue line of the numerically observed difference
starts growing, due to the development of the 
modulational instability in panels (b) and (e),
i.e., for the DNLS model, its growth appears to
be almost the same as the analytical growth rate 
$\alpha$ of the linear upper bound of Eq.~\eqref{numest}
%roughly parallel the linear slope of our 
%estimate 
(while of course resting well below
the rigorously established bound of the red line).
Lastly, when the instability growth, seeded at
the location of the localized (in space-time)
solution reaches the domain boundary, we observe
the relative difference between the two model
solutions (AL and DNLS) saturating, with the blue
line becoming nearly horizontal.
The above analysis provides a theoretical justification of the strategy followed in Section~\ref{sec_2},
namely of inserting the analytical KM solution with the prescribed set of parameters as an initial
guess for the Newton's method which resulted in finding the solution shown in Fig.~\ref{prelim_numer_2}.

\section{Branches of Numerically-Exact KM-type breather solutions to the DNLS}
% ====================================================================
The numerical results of Sec.~\ref{sec_2} paired with our theoretical
analysis on the proximity of the DNLS and AL models now raise the question
whether branches of KM-type solutions with a flat background to the DNLS
model [cf.~Eq.~\eqref{salerno}] can be obtained numerically over the background
amplitude $q$ and (breathing) frequency $\omega_{b}$. Unlike the earlier work
of~\cite{antibreathers}, where the starting point for exploration was
the anti-continuum limit, here, we leverage the existence of the KM-type
breather of Fig.~\ref{prelim_numer_2} in our numerical investigations presented
next.

In particular, we use the profile of Fig.~\ref{prelim_numer_2} with
$q=\omega_{b}=0.005$ (that our nonlinear solver converged to) as well as one 
with $q=0.005$ and $\omega_{b}=0.01$ as initial guesses to a pseudo-arclength
continuation method~\cite{kuznetsov_book_2023}, and trace branches of KM-type
breathers over the background amplitude $q$. In our computations presented in
this section, we employ a lattice with $N=200$ sites, and periodic boundary
conditions are imposed at the edges of the lattice while keeping the same number
of Fourier modes in time the same as before, i.e., $k=41$, and thus $2k+1=83$.
At each continuation step, we identify the spectral stability of the solutions
by computing the Floquet multipliers of the associated monodromy matrix, and the
stability traits of the solutions are corroborated by performing direct numerical
simulations of select profiles through the use of MATLAB's built-in \textsc{ode113}
integrator.

\begin{figure}[pt!]
\centering
\includegraphics[height=.15\textheight, angle =0]{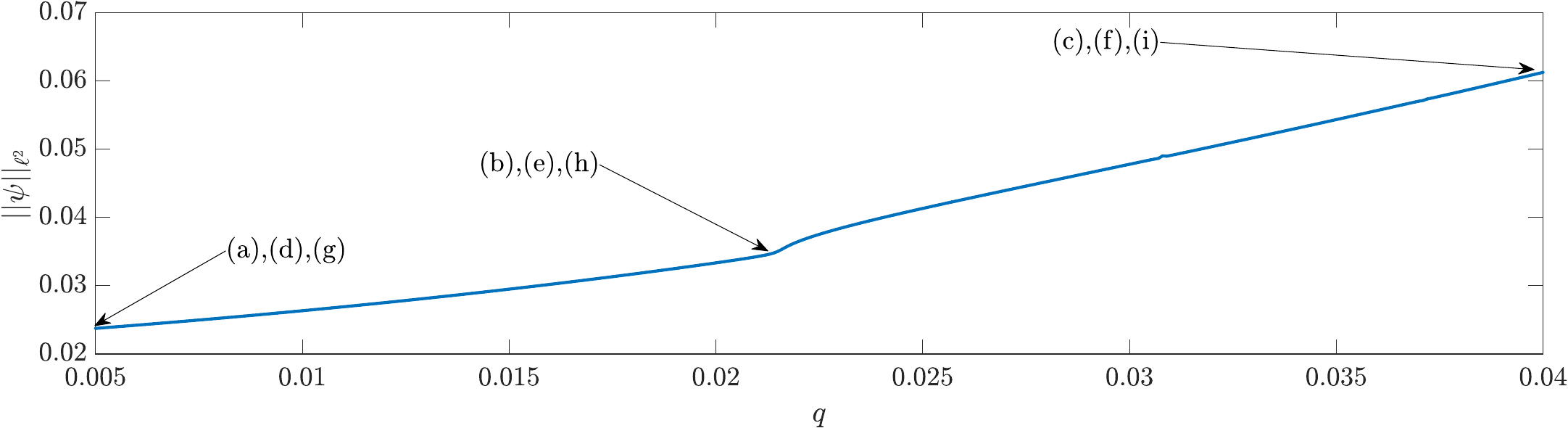}\\
\includegraphics[height=.15\textheight, angle =0]{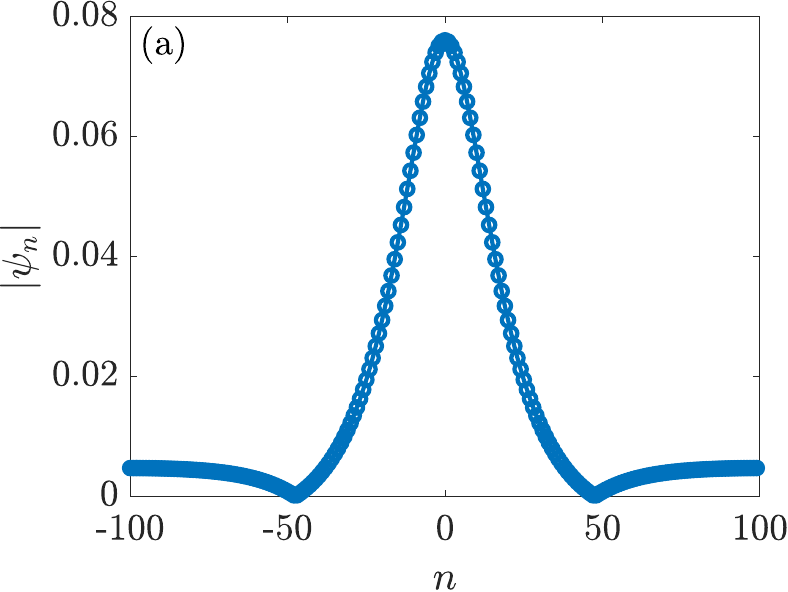}
\includegraphics[height=.15\textheight, angle =0]{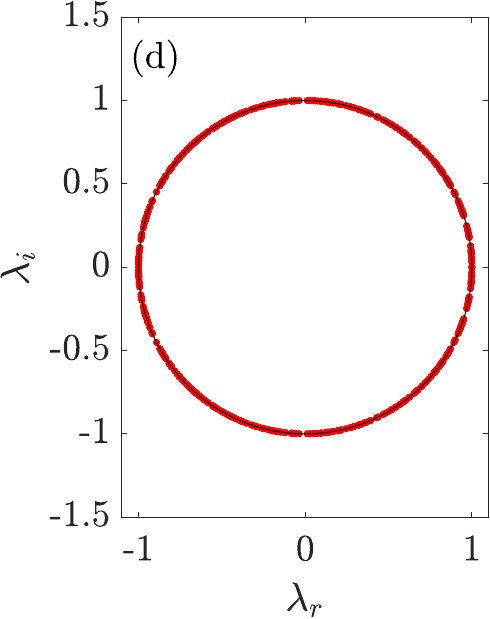}
\includegraphics[height=.15\textheight, angle =0]{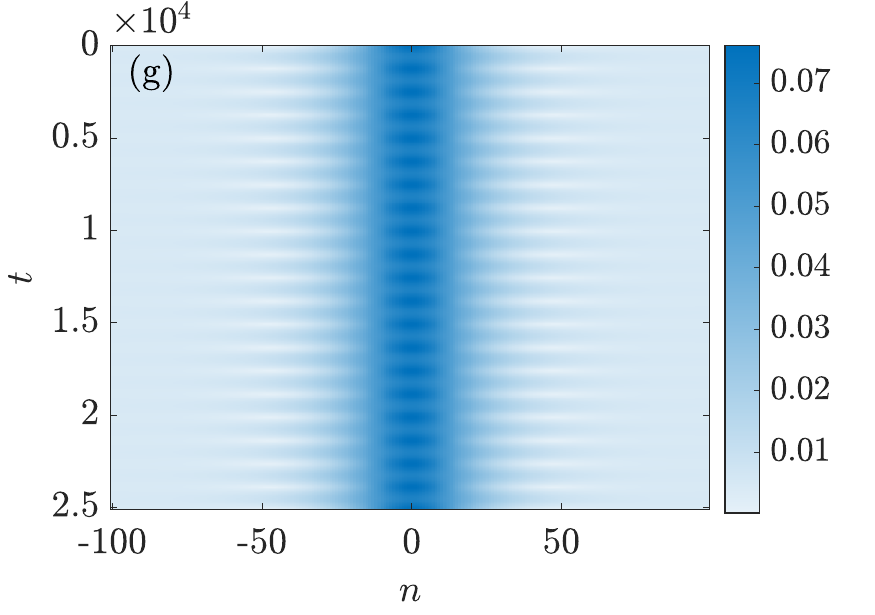}\\
\includegraphics[height=.15\textheight, angle =0]{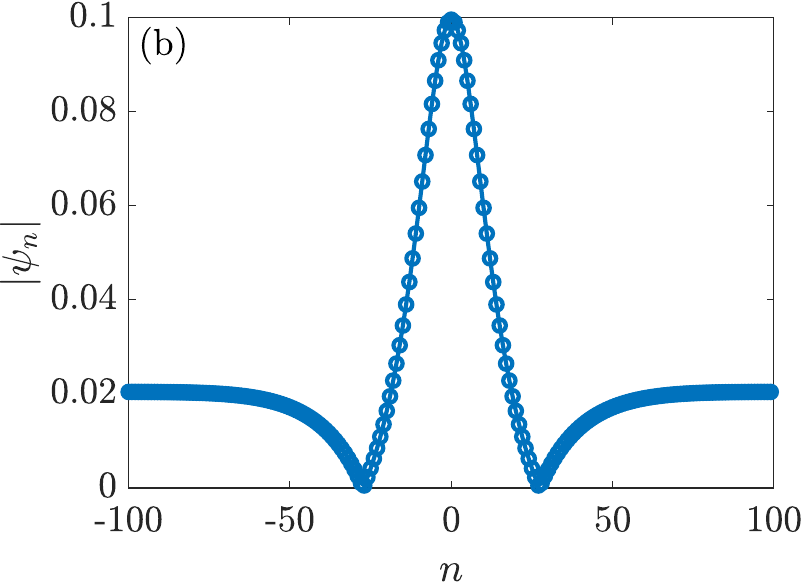}
\includegraphics[height=.15\textheight, angle =0]{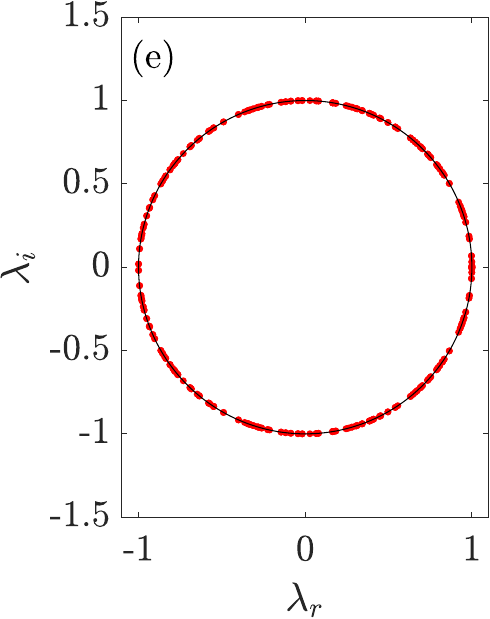}
\includegraphics[height=.15\textheight, angle =0]{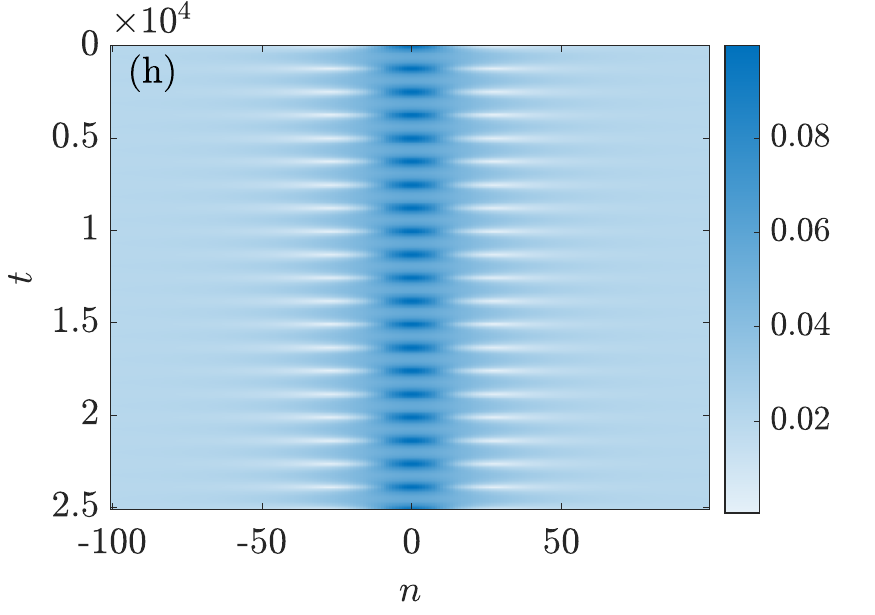}\\
\includegraphics[height=.15\textheight, angle =0]{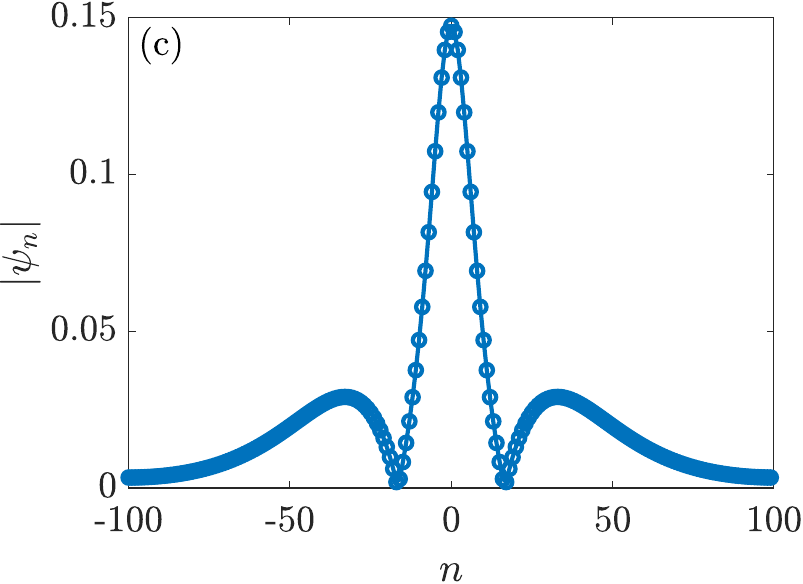}
\includegraphics[height=.15\textheight, angle =0]{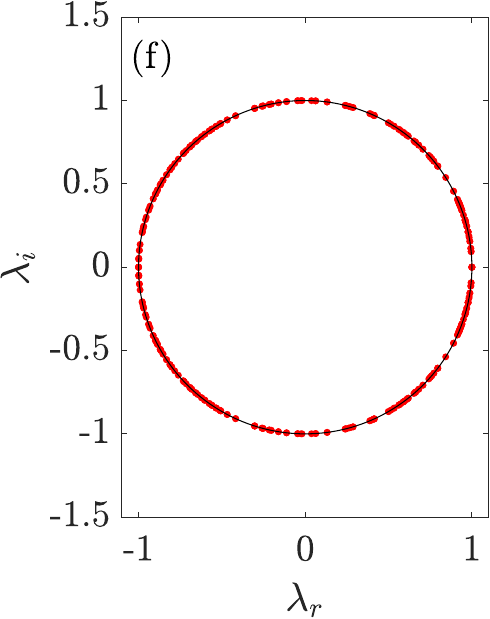}
\includegraphics[height=.15\textheight, angle =0]{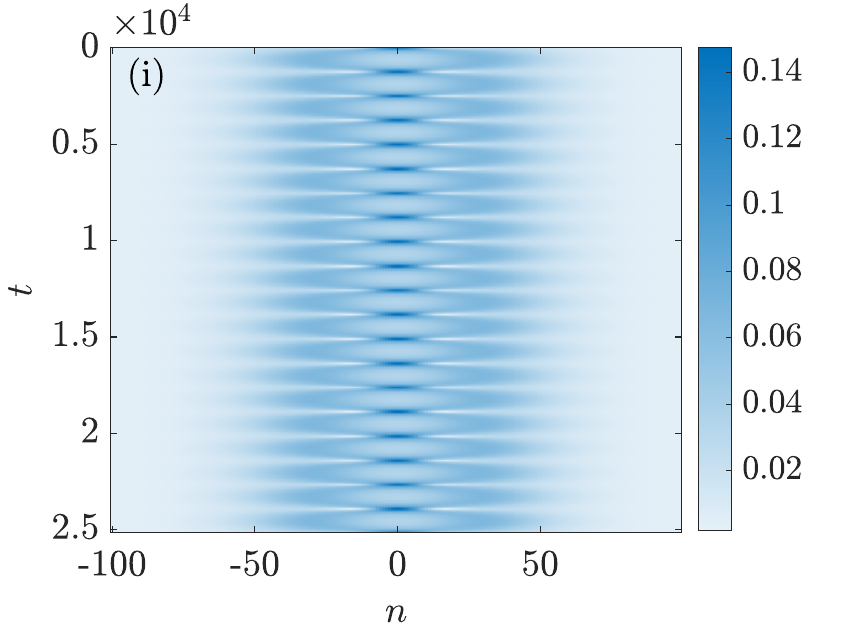}
\caption{
(Color online)
Bifurcation analysis results of KM breathers to the DNLS equation
over background amplitude $q$, and for fixed $\omega_{b} = 0.005$.
Panels (a)-(c) show numerically exact solution profiles corresponding
to values of $q=0.005$, $q=0.0214$, and $q=0.0400$, respectively. Panels
(d)-(f) give the corresponding Floquet multipliers, i.e., the eigenvalues
of the associated monodromy matrix. Finally, the panels (g)-(i) show the
dynamical evolution of the KM solutions shown in panels (a)-(c) over time,
up to $20$ periods. A switch in the phenomenology of the solutions occurs
beyond $q=0.0214$, where solutions no longer sit atop a constant background
(see, panel (c)).
\label{cont_om0.005}}
\end{figure}

\begin{figure}[pt!]
\centering
\includegraphics[height=.15\textheight, angle =0]{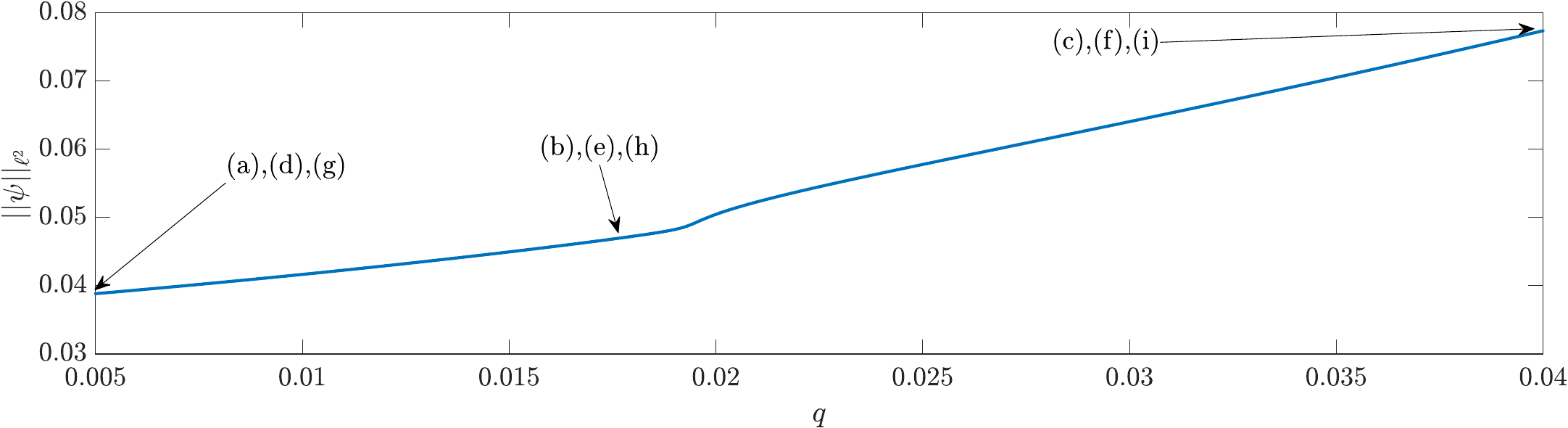}\\
\includegraphics[height=.15\textheight, angle =0]{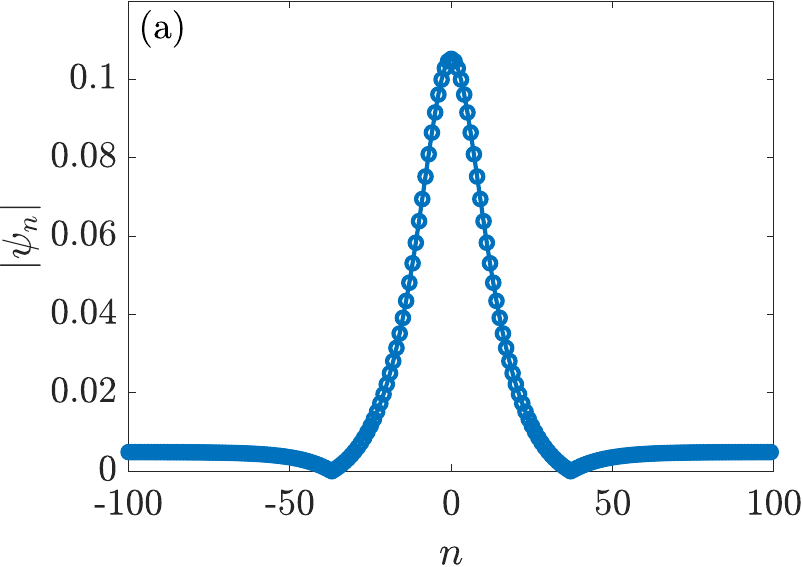}
\includegraphics[height=.15\textheight, angle =0]{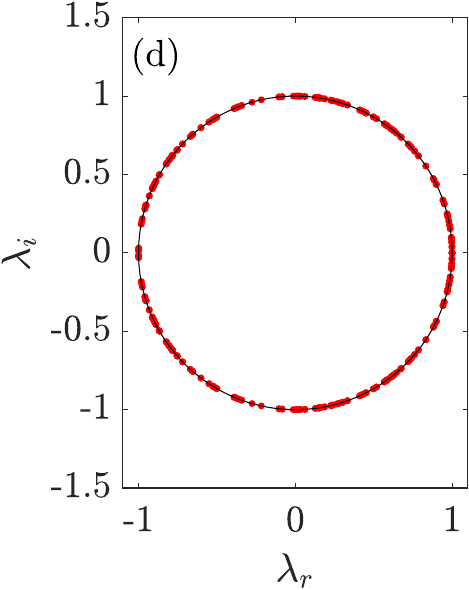}
\includegraphics[height=.15\textheight, angle =0]{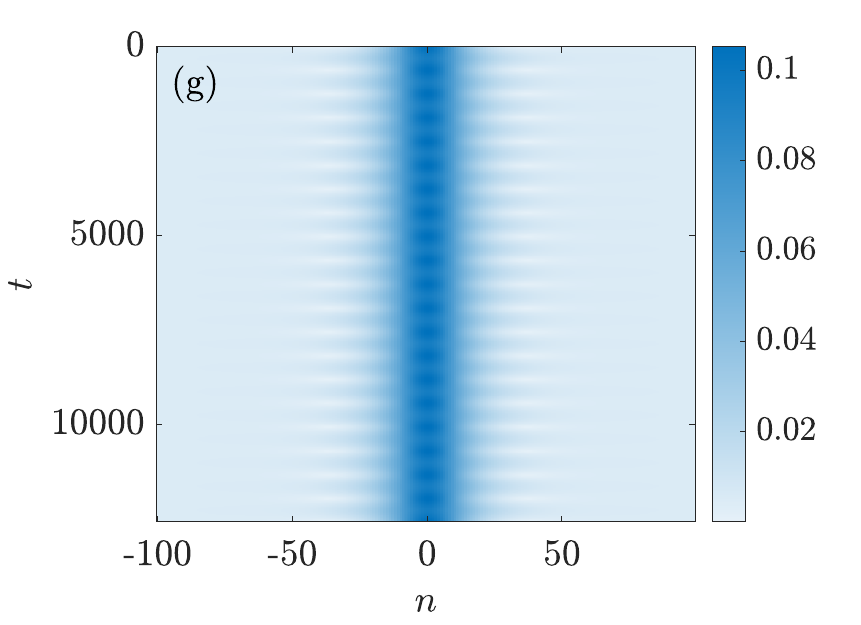}\\
\includegraphics[height=.15\textheight, angle =0]{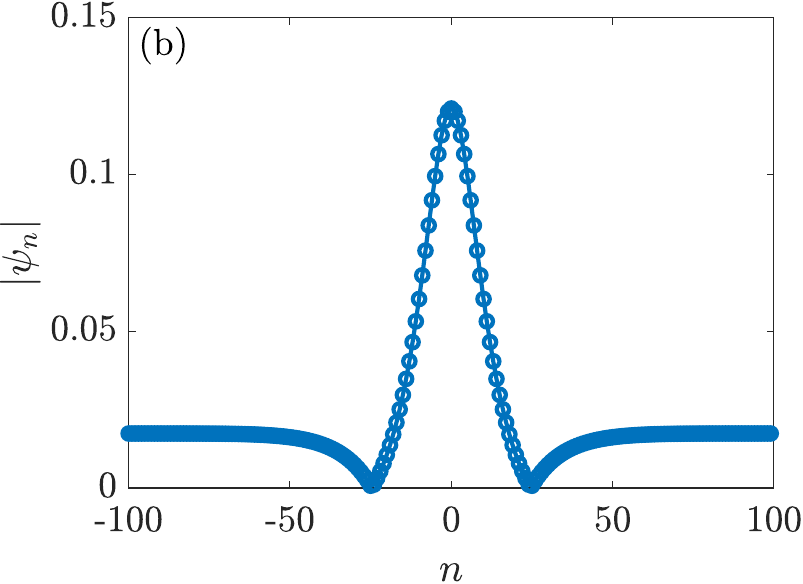}
\includegraphics[height=.15\textheight, angle =0]{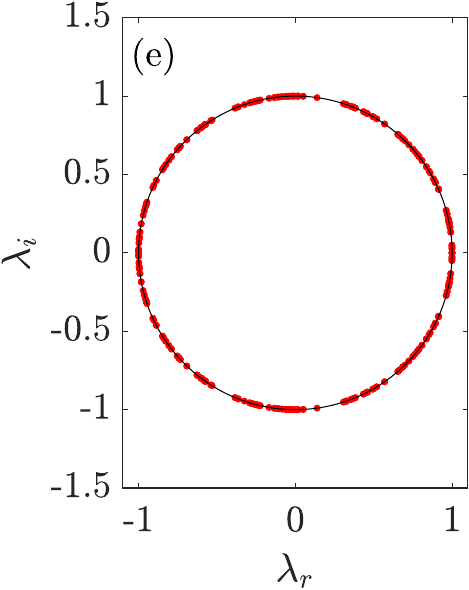}
\includegraphics[height=.15\textheight, angle =0]{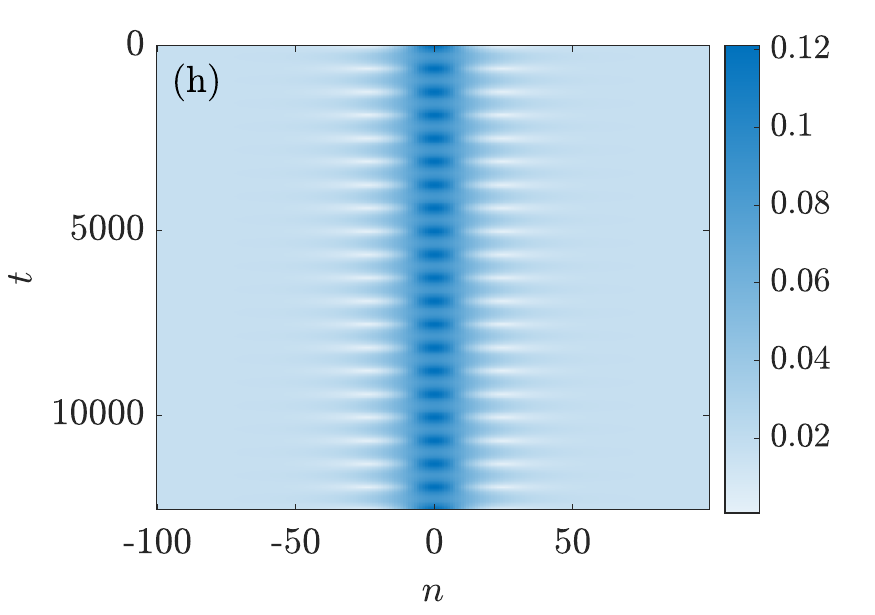}\\
\includegraphics[height=.15\textheight, angle =0]{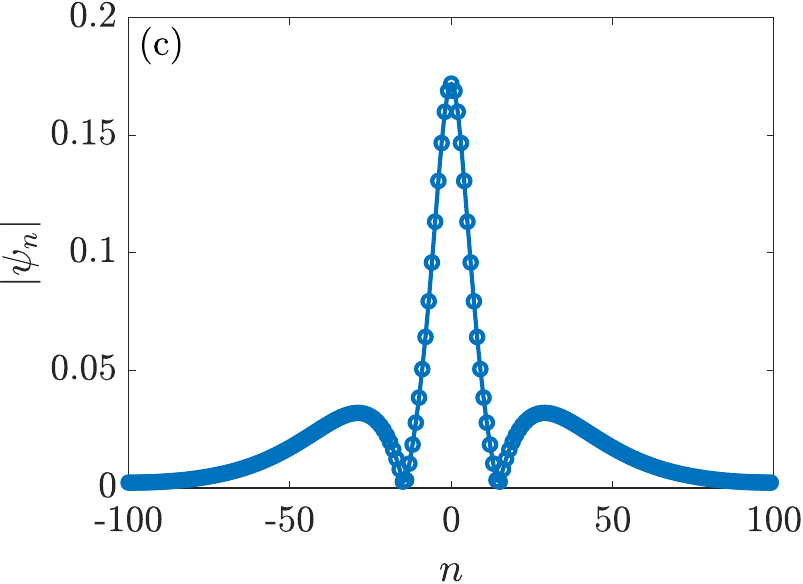}
\includegraphics[height=.15\textheight, angle =0]{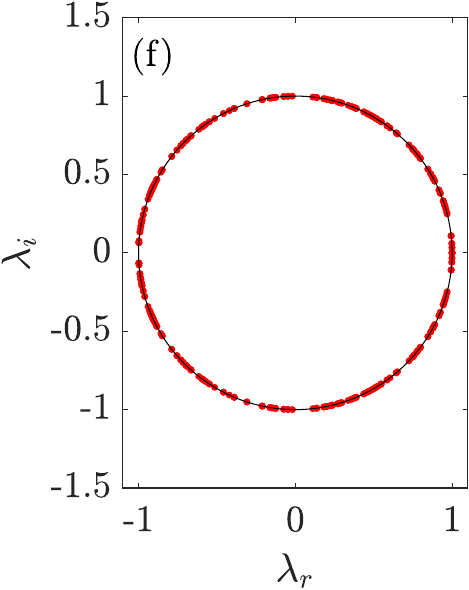}
\includegraphics[height=.15\textheight, angle =0]{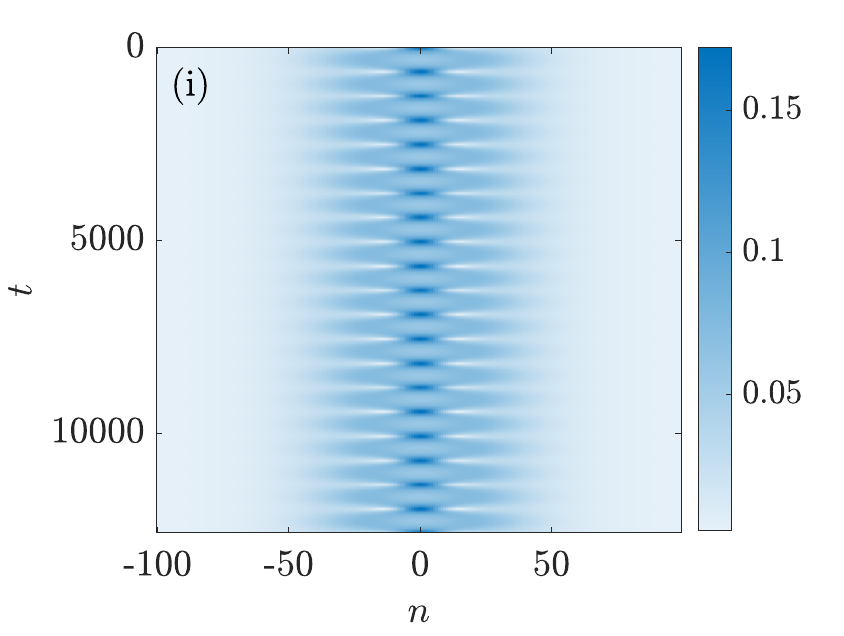}
\caption{
(Color online) Same Fig. \ref{cont_om0.005} but for $\omega_{b} = 0.01$. Panels (a)-(c)
show numerically exact solution profiles corresponding to values of $q=0.005$,
$q=0.0176$, and $q=0.0400$, respectively. Panels (d)-(f) showcase the
corresponding Floquet multipliers of these solutions. Panels (g)-(i) show the
dynamical evolution of the KM solutions shown in panels (a)-(c) over time,
(again) up to $20$ periods. A switch in the phenomenology of the solutions occurs
beyond $q=0.0176$, where solutions no longer sit atop a constant background.
\label{cont_om0.01}}
\end{figure}

We turn now our focus to the numerical results that are summarized in
Figs.~\ref{cont_om0.005} and~\ref{cont_om0.01} corresponding to KM-type
branches to the DNLS with $\omega_b=0.005$ and $\omega_b=0.01$, respectively.
The top panels in the figures showcase the dependence of the $l^{2}$ norm
of the solution on $q$, and the labels therein connect with the panels that
follow below. In particular, the left i.e., (a)-(c) and middle (d)-(f) panels
in the figures, depict the spatial distribution of the amplitude of the solutions,
i.e., $|\psi_{n}|$ and their Floquet multipliers (shown with red markers) on the
spectral plane, respectively, for different values of $q$ along the respective
branches (see, the captions in the figures for the specific values of $q$). The panels
(g)-(i) present the spatio-temporal evolution of the profiles of panels (a)-(c)
over 20 periods, i.e., $20\times T_{b}$ where $T_{b}\approx 2.5133\times10^{4}$
and $T_{b}\approx 628.3185$ in Figs.~\ref{cont_om0.005} and~\ref{cont_om0.01}, respectively.

% q   	wb	   max DNLS FM (m=200)
% 0.005	0.005	1.000841443
% 0.0214 0.005	1.000570119
% 0.04	0.005	1.000149071

% q 	wb  	max DNLS FM (m=200)
% 0.005	0.01	1.000001683
% 0.0176 0.01	1.000110454
% 0.04	0.01	1.000060798

Starting from $q=0.005$, it can be discerned from the bifurcation diagram of
Fig.~\ref{cont_om0.005} (corresponding to $\omega_{b}=0.005$) that a branch
of KM-type solutions sitting atop a constant background does exist up until
$q\approx 0.0214$, see, the profiles of panels (a) and (b), respectively, where a ``blip'' occurs that highlights the manifestation of the change in the characteristics
of the solutions. Past that value of $q$, the solutions feature same characteristics
as the KM breather of Eq.~\eqref{km_exact} but only locally, that is, the solutions
bear a peak maximum surrounded by two minima (of zero amplitude) although the spatial
asymptotics are different. Indeed, this is evident in Fig.~\ref{cont_om0.005}(c)
corresponding to $q=0.04$ where the tails of the solution's amplitude gradually
asymptote to zero for $|n|\gg 1$. A similar phenomenology is observed for the numerical
results of Fig.~\ref{cont_om0.01} corresponding to $\omega_b=0.01$.~Upon tracing the
relevant branch from $q=0.005$, the bifurcation diagram indicates the existence of
a branch of KM-type solutions atop a flat background up to $q\approx 0.0176$ whereupon
for larger values of $q$, the tails of the solution (again) asymptote to zero. This
change in the characteristics of the asymptotics of the solution is imprinted by the
presence of a blip in the bifurcation diagram shown in the top panel of Fig.~\ref{cont_om0.01}.

As far as the stability analysis results are concerned, the Floquet spectra
of panels (d)-(f) of Figs.~\ref{cont_om0.005} and~\ref{cont_om0.01} indicate
that the KM solutions to the DNLS that we report are spectrally stable via the absence of multipliers that lie off the unit circle (the
latter shown with a solid black line in the panels), i.e., $|\lambda|>1$, for
the number of lattice nodes we used.
~In line with the earlier comment made in
Sec.~\ref{sec_2}, however, all these solutions are typically expected to be MI
unstable as $N$ starts getting increased, see, Fig.~\ref{prelim_numer_3}.~Regardless,
and for the discretization we used herein, we report (after careful inspection
of the multipliers) the emergence of
%~However,
%a careful inspection of the multipliers reveals the emergence of
just one real eigenvalue
with $|\lambda|-1=\mathcal{O}(10^{-5})$ (or $\mathcal{O}(10^{-6})$) in each case we
studied with a positive real part (data not shown), a (weakly unstable) case that
is strongly reminiscent of the results of Fig.~\ref{prelim_numer_2}.%
%~This in turn suggests that the solutions are
%weakly unstable
%~Despite the presence of such a weak instability,

~Similar to Fig.~\ref{prelim_numer_2}(c), we checked the robustness of the solutions of
panels (a)-(c) of Figs.~\ref{cont_om0.005} and~\ref{cont_om0.01} by evolving them forward
in time.~The spatio-temporal dynamics shown in panels (g)-(i) give strong evidence that the
solutions breathe in time, and are indeed robust over 20 periods (note the large terminal times
of integration, which are of order of $\mathcal{O}(10^4)$).~As a further check, we performed
dynamics of the DNLS by using
perturbed versions of the KM solutions along the most unstable eigendirection (with a very small
yet positive real part) we identified in this work as initial conditions, and monitored the
breathing dynamics over $1000$ periods.~Remarkably, all the KM-type solutions that we have
identified in this work and per the number of lattice sites used are dynamically robust (data
not shown).~However, it is expected that as $N$ increases, the emergence of MI modes (see
Fig.~\ref{prelim_numer_3}) will leave an imprint in the aforementioned KM's Floquet spectra which
itself may affect the dynamics of the KM breathers, depending on the growth rates of
such unstable modes.

Based on the above results, we conclude that for
sufficiently small amplitudes of the background, 
solutions with features very proximal to those
to the AL model, that are for practical purposes
very long-lived can be found in the DNLS
model (for large lattices). 
As the value of the background increases,
the solutions change character past a ``critical
point'' (as was shown through computations for
two different frequencies), leading to waveforms
that may locally resemble the KMs of the AL
model but do not share the same asymptotic
behavior. Once again, it appears that the
smallness of the data is central to obtaining
proximal results in the form of genuine
time-periodic solutions across the DNLS and
AL model. It remains an important open question
whether such solutions can be generalized
to arbitrary amplitudes in some suitable
form in the DNLS case.

%thus suggesting that these apparent weak instabilities that appear
%in the Floquet spectrum are rather neutral eigendirections, and not true instabilities.

\section{Conclusions and Future Challenges}

In the present work we %have sought to 
explore the proximity of
the DNLS and the AL model, i.e., of the non-integrable
and integrable discretizations of the nonlinear Schr{\"o}dinger
equation. 
% This exploration has taken place at the level of
In particular, our investigation focuses on 
waveforms that are
time-periodic in the form of Kuznetsov-Ma  solitons and exist atop a finite background. These structures 
% were deemed to be
are of particular interest due to their connection with
the rogue wave nonlinear patterns. We illustrated herein
that suitable conditions (of background, frequency, etc.) can be
identified under which the DNLS and AL solutions can barely be
distinguished for long simulations in time. % under very long runs. 
This, in turn, motivated
us to explore this proximity analytically, whereby we found 
that the distance between the models grows only linearly
in time with a suitable prefactor 
%that depends on the distance between
%the initial conditions. 
depending on the size and distance between the
initial conditions in the two models.
Finally, we systematically analyzed such
waveforms and nonlinear dynamics %and their spectral stability and nonlinear dynamics
at the level of the DNLS (and AL)
models, identifying them as exact periodic solutions and
performing suitable continuations.

Naturally, this study paves the way for further explorations
along this vein.~While we have used periodic orbits herein
due to the ability to identify them in a numerically exact form
(up to prescribed tolerance), it would be relevant to
appreciate and formulate similar results for other
states, including Peregrine solitons, Akhmediev breathers,
as well as for higher order rogue waveforms. 
% Also, here, this 
Our examination also took place at the level of the discrete
models, without seeking to examine the relevant results as
a function of the discretization parameter and how the
corresponding (shared) continuum limit of the models
is approached. The latter point is also worthy of further
exploration. Finally, we restricted our considerations
to the case of focusing DNLS and AL models. However, it is
remarkable that in the case of the discrete models
such rogue patterns may arise in defocusing variants
of the models as well~\cite{Ohta_2014}, as well as the
very recent work of~\cite{Coppini_2024}. The latter
settings are also of considerable interest
at the level of the considerations presented herein.
Such studies are currently under progress and will
be reported in future publications.

\section*{Acknowledgements}
% ======================================================================
This work has been supported by the U.S. National Science Foundation
under Grants DMS-2204782 (EGC and MLL), DMS-2206270 (DM), and PHY-2110030, PHY-2408988 and DMS-2204702
(PGK). J.C.-M. acknowledges support from the EU (FEDER program 2014-2020) through
MCIN/AEI/10.13039/501100011033 under the projects PID2020-112620GB-I00 and PID2022-143120OB-I00.

\appendix
\section{Proof of Theorem \ref{Clos}}\label{app}
First, it is important to discuss some properties of the nonlinear operators appearing in equations \eqref{dbcAL} and \eqref{dbcNLS}. For the equation \eqref{dbcAL}, we consider the nonlinear operator
\begin{eqnarray}
\label{opAL1s}
\mathcal{G}_{AL}(U_n)=C(\Delta_d U)_n+C(\Delta_d q)_n+\mathcal{F}_{AL}(U_n),
\end{eqnarray}
where 
\begin{eqnarray*}
\label{opAL11}
\mathcal{F}_{AL}(U_n)=-2q^2(U_n+q_n)+|U_n+q_n|^2\left(U_{n+1}+U_{n-1}+q_{n+1}+q_{n-1}\right).
\end{eqnarray*}
The operator $\mathcal{F}_{AL}(U_n)$ can be rewritten in the form,
\begin{eqnarray}
\label{opAL2}
\mathcal{F}_{AL}(U_n)=-2q^2U_n&-&2q^2q_n+\left(|U_n|^2+\overline{q}_nU_n+q_n\overline{U}_n\right)\left(U_{n+1}+U_{n-1}+q_{n+1}+q_{n-1}\right)\nonumber\\
&+&|q_n|^2\left(U_{n+1}+U_{n-1}\right)+|q_n|^2\left(q_{n+1}+q_{n-1}\right).
\end{eqnarray}
The operator $\mathcal{G}_{AL}$ given in \eqref{opAL1s} will be well defined on bounded sets of $\ell^2$, if and only if 
\begin{eqnarray}
\label{opAL3A}
(\Delta_d q)_n=0\;\;\mbox{or}\;\; (\Delta_d q)_n\in\ell^2,
\end{eqnarray}
due to the presence of the second term of $\mathcal{G}_{AL}$, and
\begin{eqnarray}
\label{opAL3}
-2q^2q_n+|q_n|^2\left(q_{n+1}+q_{n-1}\right)=0\;\;\mbox{or}\;\; -2q^2q_n+|q_n|^2\left(q_{n+1}+q_{n-1}\right)\in\ell^2,  
\end{eqnarray}
due to the presence of the second and the last term of $\mathcal{F}_{AL}$ in \eqref{opAL2}.
By the definition of $q_n$ in \eqref{nbc2}, we have $|q_n|^2=q^2$ and, thus,  condition \eqref{opAL3} coincides with condition \eqref{opAL3A}.

Both options of \eqref{opAL3A} are valid: if $\zeta_+=\zeta_{-}$, then $(\Delta_d q)_n=0$. % \zeta = q?
If  $\zeta_+\neq\zeta_{-}$, then $(\Delta_d q)_n=(\zeta_{+}-\zeta_{-})(\delta_n-\delta_{n-1})$, where $\delta_n$ is the discrete $\delta$-function, and obviously $(\Delta_d q)_n\in\ell^2$.   

For the equation \eqref{dbcNLS}, we consider the operator 
\begin{eqnarray}
\label{opAL5}
\mathcal{G}_{DNLS}(V_n)=C(\Delta_d V)_n+C(\Delta_d q)_n+\mathcal{F}_{DNLS}(V_n),
\end{eqnarray}
where 
\begin{eqnarray*}
 \label{opAL5a} 
 \mathcal{F}_{DNLS}(V_n)=+2 \left[|V_n+q_n|^2 - q^2\right] (V_n+q_n).
\end{eqnarray*}
After some algebra, $\mathcal{F}_{DNLS}(V_n)$  is rewritten in the form
\begin{eqnarray}
\label{opAL6}
\mathcal{F}_{DNLS}(V_n)=|V_n|^2V_n+2q_n|V_n|^2+2|q_n|^2V_n
+\overline{q}_nV_n^2+q_n^2\overline{V}_n-q^2V_n+|q_n|^2q_n-q^2q_n.
\end{eqnarray}
Since $|q_n|^2=q^2$, we have 
\begin{eqnarray*}
|q_n|^2q_n-q^2q_n=0.
\end{eqnarray*}
Moreover, for the second term of the operator $\mathcal{G}_{DNLS}$, the condition \eqref{opAL3A} is satisfied as explained above.

Next, we consider the difference of solutions $D(t):=U(t)-V(t)$, of the initial-boundary value problems \eqref{dbcAL} and \eqref{dbcNLS} supplemented with the zero boundary \eqref{vbcn}, respectively. Subtracting equation \eqref{dbcNLS} from equation \eqref{dbcAL}, we see that $D(t)$ satisfies the equation
	 \begin{eqnarray}
		  \label{difD1}
		  i\dot{D_n}+C (\Delta_d D)_n=\mathcal{F}_{DNLS}(V_n)-\mathcal{F}_{AL}(U_n).
		 \end{eqnarray}

	 Multiplying \eqref{difD1} by $\overline{{D}_n}$, summing over $n$ and keeping the imaginary parts of the resulting equation, we get that $||D(t)||_{\ell^2}$ satisfies the inequality 
	 \begin{eqnarray}
		 \label{DifD2}
		 \frac{1}{2}\frac{d}{dt}||D||_{\ell^2}^2&=&\sum_{n\in\mathbb{Z}}\mathcal{F}_{DNLS}(V_n)\overline{{D}_n}-\sum_{n\in\mathbb{Z}}\mathcal{F}_{AL}(U_n)\overline{{D}_n}\nonumber\\
		 &\leq&||\mathcal{F}_{DNLS}(V)||_{\ell^2}||D||_{\ell^2}+||\mathcal{F}_{AL}(U)||_{\ell^2}||D||_{\ell^2}.
		 \end{eqnarray}
	 Since $\frac{1}{2}\frac{d}{dt}||D||_{\ell^2}^2=||D||_{\ell^2}\frac{d}{dt}||D||_{\ell^2}$, 
	 after an integration with respect to time, inequality  \eqref{DifD2} becomes
	 \begin{eqnarray}
		 \label{DifD3a}
		  ||D(t)||_{\ell^2} \leq  ||D(0)||_{\ell^2}
		  +\int_{0}^t||\mathcal{F}_{AL}(U (s))||_{\ell^2}ds+\int_{0}^t||\mathcal{F}_{DNLS}(U (s))||_{\ell^2}ds.\;\;\;\;
		 \end{eqnarray}
	 Next, we shall use the following estimates: For $||\mathcal{G}_{AL}(U (s))||_{\ell^2}$, by using the embedding 
	 \begin{equation}
		 	\label{lp2}
		 	\ell^q\subset\ell^p,\quad \|u\|_{\ell^p}\leq \|u\|_{\ell^q}, \quad 1\leq q\leq p\leq\infty,
		 \end{equation}	
	
	 for $p=4, 6$ and $q=2$, we have
	 \begin{eqnarray}
		 \label{GAL}
		 ||\mathcal{F}_{AL}(U)||_{\ell^2}&\leq& \left(\sum_{n\in\mathbb{Z}}|U_n|^4|U_{n+1}+U_{n-1}|^2\right)^{\frac{1}{2}}+2\left(\sum_{n\in\mathbb{Z}}|q_n|^2|U_n|^2|U_{n+1}+U_{n-1}|^2\right)^{\frac{1}{2}}\nonumber\\
		 &&+\left(\sum_{n\in\mathbb{Z}}|U_n|^4|q_{n+1}+q_{n-1}|^2\right)^{\frac{1}{2}}+2\left(\sum_{n\in\mathbb{Z}}|q_n|^2|U_n|^2|q_{n+1}+q_{n-1}|^2\right)^{\frac{1}{2}}\nonumber\\
		 &&+\left(\sum_{n\in\mathbb{Z}}|q_n|^4|U_{n+1}+U_{n-1}|^2\right)^{\frac{1}{2}}+2q^2\left(\sum_{n\in\mathbb{Z}}|U_n|^2\right)^{\frac{1}{2}}\nonumber\\
		 &\leq&2||U(t)||_{\ell^6}^3+6q||U(t)||^2_{\ell^4}+8q^2||U(t)||_{\ell^2}\nonumber\\
		 &\leq&2||U(t)||_{\ell^2}^3+6q||U(t)||^2_{\ell^2}+8q^2||U(t)||_{\ell^2}.
		 \end{eqnarray}
	 Similarly, for $||\mathcal{G}_{DNLS}(V (s))||_{\ell^2}$, we have the estimate
	 \begin{eqnarray}
		 \label{GDNLS}
		 ||\mathcal{F}_{DNLS}(V)||_{\ell^2}&\leq& \left(\sum_{n\in\mathbb{Z}}|V_n|^6\right)^{\frac{1}{2}}+3\left(\sum_{n\in\mathbb{Z}}|q_n|^2|V_n|^4\right)^{\frac{1}{2}}\nonumber\\
		 &&+3\left(\sum_{n\in\mathbb{Z}}|q_n|^4|V_n|^2\right)^{\frac{1}{2}}+q^2\left(\sum_{n\in\mathbb{Z}}|V_n|^2\right)^{\frac{1}{2}}\nonumber\\
		 &\leq&||V(t)||_{\ell^6}^3+3q||V(t)||^2_{\ell^4}+4q^2||V(t)||_{\ell^2}\nonumber\\
		 &\leq&||V(t)||_{\ell^2}^3+3q||V(t)||^2_{\ell^2}+4q^2||V(t)||_{\ell^2}.
		 \end{eqnarray}
	 Since $U\in C^1([0,T_{AL}^*(U(0))],\ell^2)$, that is continuously differentiable with respect to time, the assumption \eqref{sd} implies that there exist a time $\hat{T}_{AL}$, satisfying $0<\hat{T}_{AL}\leq T_{AL}^*(U(0))$, such that $||V(t)||_{\ell^2}$ will be also of order $\mathcal{O}(\varepsilon)$ for all $t\in [0, \hat{T}_{AL}]$, that is,
	 \begin{eqnarray}
		 \label{cruc1a_1}  
		 ||U(t)||_{\ell^2}\leq C_5\varepsilon,\;\;\mbox{for all}\;\;t\in [0, \hat{T}_{AL}].
		 \end{eqnarray}
	 for some $C_5>0$. Also, since $V\in C^1([0,T_{DNLS}^*(U(0)),\ell^2)$, again  the assumption \eqref{sd} implies that there exist a time $\hat{T}_{DNLS}$, satisfying $0<\hat{T}_{DNLS}\leq T_{DNLS}^*(U(0))$, such that $||V(t)||_{\ell^2}$ will be also of order $\mathcal{O}(\varepsilon)$ for all $t\in [0, \hat{T}_{DNLS}]$, that is,
	 \begin{eqnarray}
		 \label{cruc1b_1}  
		 ||V(t)||_{\ell^2}\leq C_6\varepsilon,\;\;\mbox{for all}\;\;t\in [0, \hat{T}_{DNLS}],
		 \end{eqnarray}
	 for some constant $C_6>0$. We define $T_c=\min\left\{\hat{T}_{AL},\hat{T}_{DNLS}\right\}$. Then for all $t\in[0, T_c]$, we may estimate the integral terms of the formula \eqref{DifD3a}, by using the estimates \eqref{cruc1a_1} and \eqref{cruc1b_1} in the common interval  $[0, T_c]$ for which they are both valid. This procedure will lead to the following estimate for the distance between the solutions $U(t)$ and $V(t)$
	 \begin{eqnarray}
		  \label{estdist1}   
		 ||U(t)-V(t)||_{\ell^2}&\leq& C_3\varepsilon^3t
		 +C_7(q\varepsilon^2+q^2\varepsilon)t,
		 \end{eqnarray}
	 where $C_7$ is a positive constant depending on $C_5$ and $C_6$. Inserting the assumption \eqref{sd}  on $q$, we derive the first of the estimates \eqref{closinf} with the constant $C$ depending on $C_3$ and $C_7$.  
  
  The second estimate of \eqref{closinf} is a consequence of the triangle inequality. Indeed, by using \eqref{cruc1a_1} and \eqref{cruc1b_1}, we get
  \begin{eqnarray*}
 ||U(t)-V(t)||_{\ell^2}&\leq&   ||U(t)||_{\ell^2}+ ||V(t)||_{\ell^2}\leq \hat{C}\varepsilon,\;\; \hat{C}=C_5+C_6,
  \end{eqnarray*}
  which is the second of the estimates given in \eqref{closinf}.
\bibliographystyle{apsrev4-1}
\bibliography{main.bib}

\end{document}